\documentclass[
]{jss}

\usepackage[utf8]{inputenc}

\providecommand{\tightlist}{%
  \setlength{\itemsep}{0pt}\setlength{\parskip}{0pt}}

\author{
John Paul Helveston\\George Washington University
}
\title{\pkg{logitr}: Fast Estimation of Multinomial and Mixed Logit
Models with Preference Space and Willingness to Pay Space Utility
Parameterizations}

\Plainauthor{John Paul Helveston}
\Plaintitle{logitr: Fast Estimation of Multinomial and Mixed Logit
Models with Preference Space and Willingness to Pay Space Utility
Parameterizations}
\Shorttitle{\pkg{logitr}: Preference and WTP Space Logit Models in
\proglang{R}}

\Abstract{
This paper introduces the \pkg{logitr} \proglang{R} package for fast
maximum likelihood estimation of multinomial logit and mixed logit
models with unobserved heterogeneity across individuals, which is
modeled by allowing parameters to vary randomly over individuals
according to a chosen distribution. The package is faster than other
similar packages such as \pkg{mlogit}, \pkg{gmnl}, \pkg{mixl}, and
\pkg{apollo}, and it supports utility models specified with ``preference
space'' or ``willingness to pay (WTP) space'' parameterizations,
allowing for the direct estimation of marginal WTP. The typical
procedure of computing WTP post-estimation using a preference space
model can lead to unreasonable distributions of WTP across the
population in mixed logit models. The paper provides a discussion of
some of the implications of each utility parameterization for WTP
estimates. It also highlights some of the design features that enable
\pkg{logitr}'s performant estimation speed and includes a benchmarking
exercise with similar packages. Finally, the paper highlights additional
features that are designed specifically for WTP space models, including
a consistent user interface for specifying models in either space and a
parallelized multi-start optimization loop, which is particularly useful
for searching the solution space for different local minima when
estimating models with non-convex log-likelihood functions.
}

\Plainkeywords{logit, utility, preference, willingness to pay, discrete
choice models, R, maximum likelihood estimation}

\Address{
    John Paul Helveston\\
    George Washington University\\
    Department of Engineering Management and Systems Engineering,\\
School of Engineering and Applied Sciences,\\
800 22nd St NW\\
Washington, DC 20052\\
  E-mail: \email{jph@gwu.edu}\\
  URL: \url{https://www.jhelvy.com}\\~\\
  }

\usepackage{amsmath}
\usepackage{booktabs}
\usepackage{caption}
\usepackage{longtable}

\usepackage{upgreek} \usepackage{booktabs} \usepackage{float} \usepackage{array} \usepackage{booktabs} \usepackage[margin=1in]{geometry} \usepackage{graphicx} \usepackage{amsmath} \usepackage[margin=1cm, font=normalsize]{caption}

\begin{document}

\newcommand{\muVec}{\boldsymbol\upmu}
\newcommand{\sigmaVec}{\boldsymbol\upsigma}
\newcommand{\betaVec}{\boldsymbol\upbeta}
\newcommand{\thetaVec}{\boldsymbol\uptheta}
\newcommand{\omegaVec}{\boldsymbol\upomega}
\newcommand{\zetaVec}{\boldsymbol\upzeta}
\newcommand{\deltaVec}{\boldsymbol\updelta}
\newcommand{\gammaVec}{\boldsymbol\upgamma}
\newcommand{\epsilonVec}{\boldsymbol\upepsilon}
\newcommand{\xVec}{\mathrm{\mathbf{x}}}
\newcommand{\XVec}{\mathrm{\mathbf{X}}}

\section[intro]{Introduction}\label{intro}

Choice modeling is a well-established statistical method for assessing
consumer preferences across a wide variety of fields. One of the most
common approaches for modeling choice is the maximum likelihood
estimation of multinomial logit models \citep{McFadden1974}, which is
rooted in the theory of random utility models
\citep{Louviere2000, Train2009}. The central assumption of these models
is that individual consumers make choices that maximize an underlying
random utility model, which can be parameterized as a function of a
product's observed attributes and a random variable representing the
portion of utility unobservable to the modeler. These models produce
estimates of the marginal utility for changes in each attribute relative
to one another.

In many applications, modelers are interested in estimating marginal
``willingness to pay'' (WTP) for changes in product attributes. The
typical procedure to obtain these estimates is to divide the estimated
parameters of a ``preference space'' utility model by the negative of
the price parameter. Despite this common practice, it can yield
unreasonable distributions of WTP across the population in heterogeneous
random parameter (or ``mixed logit'') models
\citep{Train2005, Sonnier2007, Helveston2018}. For example, if the
parameters for the price attribute and another non-price attribute are
both assumed to be normally distributed across the population, then the
resulting WTP estimate follows a Cauchy distribution, implying that WTP
has an infinite variance across the population.

An alternative approach is to re-parameterize the utility model into the
``WTP space'' prior to estimation. Estimating a WTP space model allows
the modeler to directly specify assumptions of how WTP is distributed,
which has been found to yield more reasonable estimates of WTP
\citep{Train2005, Sonnier2007, Daly2012}. WTP space models have also
been found to be more consistent with respondent's true underlying
preferences \citep{Crastesa2014}. Finally, since WTP estimates are
independent of error scaling, they can be conveniently compared across
different models estimated on different data.

Several statistical packages support the estimation of multinomial and
mixed logit models with WTP space utility parameterizations. One of the
most common approaches involves an adaptation of the generalized
multinomial logit (GMNL) model \citep{Fiebig2010} to fit WTP space
models via an implementation of the scaled multinomial logit (SMNL)
model, though this requires that the price parameter estimate and
standard error be calculated post-estimation. Estimation of WTP space
models via GMNL has been implemented in R with the \pkg{gmnl} package
\citep{Sarrias2017} and in \proglang{STATA} with the \pkg{gmnl} package
\citep{Gu2013}. WTP space models can also be estimated using the
\pkg{apollo} \citep{Hess2019} and \pkg{mixl} \citep{Molloy2021} R
packages as they allow the user to hand-specify any valid utility model.
Finally, Professor Arne Rise Hole developed two \proglang{STATA}
packages that share a common syntax for estimating mixed logit models in
the preference space (\pkg{mixlogit}) and WTP space (\pkg{mixlogitwtp})
\citep{Hole2007}. Many other packages exist for estimating a wider
variety of logit models, but they are limited to preference space
models. Of these, package \pkg{mlogit} \citep{Croissant2020} is perhaps
the most complete and widely used for estimating multinomial logit and
mixed logit models in R via maximum likelihood estimation.

The \pkg{logitr} package is designed specifically to support the
estimation of multinomial logit and mixed logit models models with
either preference space or WTP space utility parameterizations. While
\pkg{logitr} is less general in scope compared to more flexible packages
like \pkg{mixl} and \pkg{apollo}, it offers other functionality that is
particularly useful for estimating WTP space models and conveniently
switching between preference and WTP space models. For example, given
their non-linear utility specification, WTP space models often diverge
during estimation and can be sensitive to starting parameters. To
address this, the package includes a parallelized multi-start
optimization loop to search for different local minima from different
random starting points when minimizing the negative log-likelihood. The
user interface is also more streamlined and simplified for estimating
models in either space.

Package \pkg{logitr} is also computationally efficient and faster than
other similar packages, including the \pkg{mixl} package which uses high
performance C++ code to compile the log-likelihood function
\citep{Molloy2021} (though \pkg{mixl} can be accelerated considerably
via multi-core processing). The performance gains are the result of a
combination of design features, including how the choice probabilities
are specified, avoiding redundant computation by pre-computing constant
intermediate variables, and the use of analytic gradients that are
optimized for efficiency.

The rest of the article is organized as follows: Section 2 provides an
overview of the models supported by \pkg{logitr}, including multinomial
and mixed logit models with preference space and WTP space utility
parameterizations. Section 3 discusses several important implications of
preference versus WTP space utility parameterizations on WTP estimates.
Section 4 describes the software architecture and performance. Section 5
then introduces the \pkg{logitr} package, including examples of
estimating multinomial and mixed logit models in both preference and WTP
spaces as well as additional functionality for estimating weighted
models and making predictions. Section 6 explains some limitations of
WTP space models. Finally, Section 7 concludes the paper.

\hypertarget{models}{%
\section{Models}\label{models}}

\hypertarget{the-random-utility-model-in-two-spaces}{%
\subsection{The random utility model in two
spaces}\label{the-random-utility-model-in-two-spaces}}

Random utility models assume that consumers choose the alternative \(j\)
from a set of alternatives that has the greatest utility \(u_{j}\).
Utility is a random variable that is modeled as
\(u_{j} = v_{j} + \varepsilon_{j}\), where \(v_{j}\) is the ``observed
utility'' (a function of the observed attributes such that
\(v_{j} = f(\mathrm{\mathbf{x}}_{j})\)) and \(\varepsilon_{j}\) is a
random variable representing the portion of utility unobservable to the
modeler.

Adopting the same notation as in Helveston et al.
\citeyearpar{Helveston2018}, consider the following utility model:

\begin{equation}
    u^*_{j} =
        \boldsymbol\upbeta^{*\top} \mathrm{\mathbf{x}}_{j} +
        \alpha^{*} p_{j} +
        \varepsilon^{*}_{j},
        \quad\quad
        \varepsilon^{*}_{j} \sim \textrm{Gumbel}\left(0, \sigma^2\frac{\pi^2}{6}\right)
\label{eq:utility}
\end{equation}

\noindent where \(\boldsymbol\upbeta^{*}\) is the vector of coefficients
for non-price attributes \(\mathrm{\mathbf{x}}_{j}\), \(\alpha^{*}\) is
the coefficient for price \(p_{j}\), and the error term,
\(\varepsilon^{*}_{j}\), is an IID random variable with a Gumbel extreme
value distribution of mean zero and variance \(\sigma^2(\pi^2/6)\).

This model is not identified since there exists an infinite set of
combinations of values for \(\boldsymbol\upbeta^{*}\), \(\alpha^{*}\),
and \(\sigma\) that will produce the same choice probabilities. In order
to specify an identifiable model, Equation \ref{eq:utility} must be
normalized. One approach is to normalize the scale of the error term by
dividing Equation \ref{eq:utility} by \(\sigma\), producing the
``preference space'' utility specification \citep{Train2005}:

\begin{equation}
    \left(\frac{u^*_{j}}{\sigma}\right) =
        \left( \frac{\boldsymbol\upbeta^{*}}{\sigma} \right)^\top \mathrm{\mathbf{x}}_{j} +
        \left( \frac{\alpha^{*}}{\sigma} \right) p_{j} +
        \left( \frac{\varepsilon^{*}_{j}}{\sigma} \right),
        \quad\quad
        \left( \frac{\varepsilon^{*}_{j}}{\sigma} \right) \sim \textrm{Gumbel}\left(0, \frac{\pi^2}{6}\right)
\label{eq:utilityPreferenceScaled}
\end{equation}

\noindent The typical preference space parameterization of the
multinomial logit model can then be written by rewriting Equation
\ref{eq:utilityPreferenceScaled} with \(u_j = (u^*_j / \sigma)\),
\(\boldsymbol\upbeta= (\boldsymbol\upbeta^{*} / \sigma)\),
\(\alpha = (\alpha^{*} / \sigma)\), and
\(\varepsilon_{j} = (\varepsilon^{*}_{j} / \sigma)\):

\begin{equation}
    u_{j} =
        \boldsymbol\upbeta^\top \mathrm{\mathbf{x}}_{j} +
        \alpha p_{j} +
        \varepsilon_{j}
        \hspace{0.5in}
        \varepsilon_{j} \sim \textrm{Gumbel}\left(0,\frac{\pi^2}{6}\right)
\label{eq:utilityPreference}
\end{equation}

The vector \(\boldsymbol\upbeta\) in Equation \ref{eq:utilityPreference}
represents the marginal utility for changes in each non-price attribute
(relative to the standardized scale of the error term), and \(\alpha\)
represents the marginal utility obtained from changes in price (relative
to the standardized scale of the error term). The coefficients
\(\boldsymbol\upbeta\) and \(\alpha\) are only relative values rather
than absolute and do not have units. Using this model, estimates of the
marginal WTP for changes in each non-price attribute could be computed
by dividing \(\hat{\boldsymbol\upbeta}\) by \(- \hat{\alpha}\), where
the ``hat'' symbol indicates a parameter estimate.

An alternative approach to normalizing Equation \ref{eq:utility} is to
divide by \(- \alpha^*\) instead of \(\sigma\), resulting in the ``WTP
space'' utility parameterization:

\begin{equation}
    \left(\frac{u^*_{j}}{- \alpha^*}\right) =
        \left(\frac{\boldsymbol\upbeta^{*}}{- \alpha^{*}}\right)^\top \mathrm{\mathbf{x}}_{j} +
        \left(\frac{\alpha^{*}}{- \alpha^{*}}\right) p_{j} +
        \left(\frac{\varepsilon^{*}_{j}}{- \alpha^{*}}\right),
        \quad\quad
        \left(\frac{\varepsilon^{*}_{j}}{- \alpha^{*}}\right) \sim \textrm{Gumbel} \left(0, \frac{\sigma^2}{(- \alpha^{*})^2}\frac{\pi^2}{6} \right)
\label{eq:utilityWtpScaled}
\end{equation}

Since the error term in Equation \ref{eq:utilityWtpScaled} is scaled by
\(\lambda^2 = \sigma^2/(- \alpha^{*})^2\), it can be rewritten by
multiplying both sides by \(\lambda= (- \alpha^{*} / \sigma\)) and
renaming \(u_j = (\lambda u^*_j / - \alpha^*)\),
\(\boldsymbol\upomega= (\boldsymbol\upbeta^{*} / - \alpha^{*}\)), and
\(\varepsilon_j = (\lambda \varepsilon^*_j / - \alpha^*)\):

\begin{equation}
    u_{j} =
        \lambda \left(
            \boldsymbol\upomega^\top \mathrm{\mathbf{x}}_{j} - p_{j}
            \right) +
        \varepsilon_{j}
        \hspace{0.5in}
        \varepsilon_{j} \sim \textrm{Gumbel}\left(0, \frac{\pi^2}{6}\right)
\label{eq:utilityWtp}
\end{equation}

The vector \(\boldsymbol\upomega\) in Equation \ref{eq:utilityWtp}
represents the marginal WTP for changes in each non-price attribute, and
\(\lambda\) represents the scale of the deterministic portion of utility
relative to the standardized scale of the random error term (also called
the ``scale parameter''). In contrast to the \(\boldsymbol\upbeta\)
coefficients from the preference space model in Equation
\ref{eq:utilityPreference}, the \(\boldsymbol\upomega\) coefficients
have absolute value with units of currency.

The \pkg{logitr} package can fit logit models with either utility
parameterization, and it contains functions that facilitate the
comparison of WTP estimates between models from the two model spaces.

\hypertarget{multinomial-and-mixed-logit-probabilities}{%
\subsection{Multinomial and mixed logit
probabilities}\label{multinomial-and-mixed-logit-probabilities}}

By assuming that the error term in Equations \ref{eq:utilityPreference}
and \ref{eq:utilityWtp} follows a Gumbel extreme value distribution, the
probability that a consumer will choose alternative \(j\) in choice
situation \(n\) follows a convenient, closed form expression, cf.~Train
\citeyearpar{Train2009}:

\begin{equation}
    P_{nj} =
        \frac{
            \exp \left( v_{nj} \right)
        }
        {
            \sum_k^J
            \exp \left( v_{nk} \right)
        },
    \label{eq:logitProbabilityMnl}
\end{equation}

\noindent  where \(v_{nj}\) is the deterministic portion of the utility
model and \(J\) is the number of alternatives in choice situation \(n\).
The multinomial logit model assumes homogeneous preferences across the
population and possess the independence of irrelevant alternatives (IIA)
property, which means that the ratio of any two probabilities is
independent of the functions determining any other outcome since

\begin{equation}
    \frac{P_{nj}}{P_{nk}} =
        \frac{
            \exp \left( v_{nj} \right)
        }
        {
            \exp \left( v_{nk} \right)
        },
    \label{eq:iia}
\end{equation}

To relax this assumption and allow for heterogeneity of preferences
across the population, the multinomial logit model can be extended to
the random coefficients ``mixed'' logit model \citep{McFadden2000} where
probabilities are the integrals of standard logit probabilities over a
density of parameters across people:

\begin{equation}
    P_{nj} =
    \int \left (
        \frac{
            \exp \left( v_{nj} \right)
        }
        {
            \sum_k^J
            \exp \left( v_{nk} \right)
        }
        \right )
        f(\boldsymbol\uptheta) d \boldsymbol\uptheta,
    \label{eq:logitProbabilityMxl}
\end{equation}

\noindent  where \(f(\boldsymbol\uptheta)\) is a density function and
\(\boldsymbol\uptheta\) contains the parameters in the deterministic
portion of the utility model, which are \(\boldsymbol\upbeta\) and
\(\alpha\) for preference space models (Equation
\ref{eq:utilityPreference}) and \(\boldsymbol\upomega\) and \(\lambda\)
for WTP space models (Equation \ref{eq:utilityWtp}). The mixed logit
probability can be interpreted as a weighted average of the multinomial
logit probability with weights given by the density
\(f(\boldsymbol\uptheta)\).

Modelers often specify different mixing distributions for parameters in
\(\boldsymbol\uptheta\) depending on assumptions of how preferences
might be distributed across the population. For example, modelers may
assume \(\alpha\) follows a log-normal or zero-censored normal
distribution to force the price coefficient to remain positive---an
assumption based on the logic that most people prefer price
\emph{decreases} rather than increases. Likewise, parameters in
\(\boldsymbol\upbeta\) are often assumed to follow a normal distribution
if it is unclear whether the utility parameters for attributes
\(\mathrm{\mathbf{x}}_j\) should be positive or negative.

\hypertarget{maximum-likelihood-estimation}{%
\subsection{Maximum likelihood
estimation}\label{maximum-likelihood-estimation}}

Parameters in the preference or WTP space utility models can be
estimated by maximizing the log-likelihood function. For the multinomial
logit model, the log-likelihood is given by:

\begin{equation}
    L =
  \sum_n^N
  \sum_j^J
    y_{nj} \ln P_{nj}
    \label{eq:logLikelihood}
\end{equation}

\noindent  where \(y_{nj} = 1\) if alternative \(j\) is chosen in
situation \(n\) and \(0\) otherwise, \(N\) is the number of choice
situations, \(J\) is the number of alternatives in choice situation
\(n\), and the probabilities \(P_{nj}\) are given by Equation
\ref{eq:logitProbabilityMnl}.

For mixed logit models, the log-likelihood can be estimated using
simulation to obtain estimates of \(P_{nj}\) in Equation
\ref{eq:logitProbabilityMxl} \citep{Train2009}. Over a series of
iterations, parameters are drawn from \(f(\boldsymbol\uptheta)\) and
used to compute the logit probability in Equation
\ref{eq:logitProbabilityMnl}. The average probabilities over all of the
iterations, \(\hat{P}_{nj}\), are then used in place of \(P_{nj}\) in
Equation \ref{eq:logLikelihood} to compute the \emph{simulated}
log-likelihood. Should the data contain a panel structure where multiple
observations come from the same individual, the product of the logit
probabilities in Equation \ref{eq:logitProbabilityMnl} over all trials
for each individual must first computed and then averaged over the draws
of each parameter drawn from \(f(\boldsymbol\uptheta)\)
\citep{Train2009}.

McFadden \citeyearpar{McFadden1974} shows that the log-likelihood
function is globally concave for linear-in-parameters utility models
with fixed parameters. This implies that optimization algorithms should
always arrive at a global solution when minimizing the negative
log-likelihood for preference space models with fixed parameters. In
contrast, WTP space utility models (as well as mixed logit models with
either utility parameterization) have non-convex log-likelihood
functions and thus are not guaranteed to arrive at a global solution.
For these models, different optimization strategies should be used to
minimize the negative log-likelihood, such as using a multi-start loop
where the optimization algorithm is run multiple times from different
random starting points to search for multiple local minima.

\hypertarget{implications-of-preference-versus-wtp-space-utility-parameterizations}{%
\section{Implications of preference versus WTP space utility
parameterizations}\label{implications-of-preference-versus-wtp-space-utility-parameterizations}}

WTP estimates can be obtained from both preference and WTP space utility
parameterizations. In the preference space utility model given by
Equation \ref{eq:utilityPreference}, WTP is estimated as
\(\hat{\boldsymbol\upbeta} / - \hat{\alpha}\); in the WTP space model
given by Equation \ref{eq:utilityWtp}, WTP is simply
\(\hat{\boldsymbol\upomega}\). The choice of which approach to use can
have important implications for estimates of WTP, and modelers should
consider which outcomes and measures are most relevant to any one
particular study when making a choice between the two parameterizations.

\hypertarget{distribution-of-wtp-estimates-across-the-population}{%
\subsection{Distribution of WTP estimates across the
population}\label{distribution-of-wtp-estimates-across-the-population}}

Depending on the utility parameterization used, the distribution of WTP
in mixed logit models can be sensitive to distributional assumptions of
model parameters \citep{Train2005, Sonnier2007}. For example, in a
preference space model, if \(\alpha\) and \(\beta\) were each assumed to
be normally distributed, then the WTP for marginal changes in \(x_j\)
would follow a Cauchy distribution, implying that WTP has an infinite
variance across the population. This WTP distribution is not likely what
the modeler had in mind when making individual distributional
assumptions on \(\alpha\) and \(\beta\), but it is the implied result.
In contrast, in a WTP space model the distribution of WTP for marginal
changes in \(x_j\) can be directly specified.

Several prior studies have also identified this issue, and all find that
WTP space utility parameterizations yield more reasonable estimates of
WTP. In a study on preferences for alternative-fuel vehicles, Train and
Weeks \citeyearpar{Train2005} found that while a preference space model
with a log-normally distributed price coefficient fit the data better,
it resulted in unreasonably large estimates of WTP; in contrast, a WTP
space model produced much more reasonable estimates of WTP. Using a
Bayesian approach, Sonnier et al. \citeyearpar{Sonnier2007} similarly
found that a preference space model with heterogeneity distributions for
attribute and price coefficients resulted in poorly behaved posterior
WTP distributions and that the problem was particularly bad in small
sample settings. Finally, Daly et al. \citeyearpar{Daly2012} show that
when the price coefficient is modeled with a variety of popular
distributions, including the normal, truncated normal, uniform, and
triangular, the resulting distribution of WTP has infinite moments.

To illustrate this issue, consider an example of three preference space
models with different assumptions on how the price parameter is
distributed. This example uses the \texttt{yogurt} data set included in
package \pkg{logitr}. In each model, coefficients for the yogurt brand
(\(\boldsymbol\upbeta\) in the preference space and
\(\boldsymbol\upomega\) in the WTP space) are modeled as normally
distributed. However, the price parameter, \(\alpha\), in the preference
space model (and likewise the scale parameter, \(\lambda\), in the WTP
space model) is modeled three different ways: 1) as a fixed coefficient,
2) normally distributed, and 3) log-normally distributed.

\begin{figure}[h]
{\centering \includegraphics[width=1\linewidth]{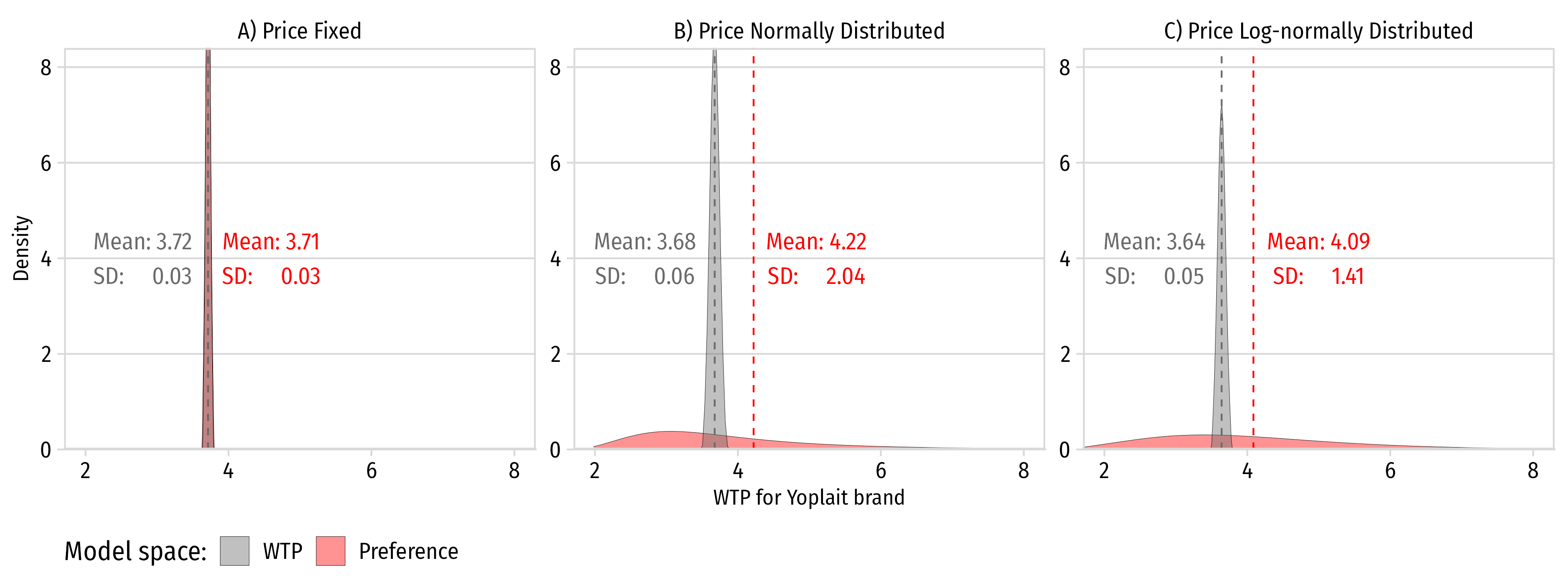}}
\caption[Comparison of WTP distribution for the Yoplait brand from mixed logit models with preference space (red) and WTP space (gray) utility parameterizations]{Comparison of WTP distribution for the Yoplait brand from mixed logit models with preference space (red) and WTP space (gray) utility parameterizations. In each panel, the price parameter is modeled as fixed, normally distributed, or log-normally distributed. The ``SD'' in the labels means ``Standard Deviation''.}
\label{fig:fig1}
\end{figure}

Figure \ref{fig:fig1} compares the WTP distribution for the Yoplait
brand across the population from each preference space and WTP space
model. In the case where \(\alpha\) and \(\lambda\) are modeled as fixed
coefficients (panel A), the WTP distributions from each model are
identical. But when \(\alpha\) and \(\lambda\) are modeled as normally
distributed (panel B), the WTP distribution from the preference space
model changes dramatically. Since WTP in this case is the ratio of two
normal distributions, the variance increases by two orders of magnitude
and the mean shifts upward. A similar outcome occurs when \(\alpha\) and
\(\lambda\) are modeled as log-normally
distributed\footnote{Log-normal distributions are often used to force parameter positivity.}
(panel C). Note that the WTP distribution in the WTP space model remains
nearly the same regardless of how \(\lambda\) was assumed to be
distributed.

This example illustrates the sensitivity of the WTP distribution to
modeling choices made in preference space utility models. In general,
WTP space models yield more reasonable estimates of WTP distributions
across the population, a consistent finding across multiple prior
studies \citep{Train2005, Sonnier2007, Das2009, Helveston2018}.

\hypertarget{prediction}{%
\subsection{Prediction}\label{prediction}}

Whether preference space or WTP space utility models predict better is
an empirical question that has not been definitively addressed in prior
studies. Of the few studies that have estimated models in the WTP space,
none have provided conclusive evidence that one parameterization
systematically predicts better than the other. Both Sonnier et al.
\citeyearpar{Sonnier2007} and Train and Weeks \citeyearpar{Train2005}
found that a preference space parameterization fit the data better, but
they also both found that the resulting estimates of WTP had
unreasonably large tails. Das et al. \citeyearpar{Das2009} also
estimated both preference and WTP space models using data on preferences
for landfill site attributes in Rhode Island, and they found nearly
identical model fit on out-of-sample predictions with each model
specification, though the WTP space model yielded more reasonable
estimates of WTP.

\hypertarget{practical-considerations}{%
\subsection{Practical considerations}\label{practical-considerations}}

In addition to considerations of WTP estimates, model fit, and
prediction, there are other practical implications to consider when
choosing to estimate a preference versus WTP space model. Perhaps the
simplest and most obvious distinction is that a WTP space model yields
estimates of WTP without needing additional post-estimation
calculations. And because WTP space coefficients have units of currency,
they have a concrete meaning that can be immediately interpreted. In
contrast, preference space model coefficients only have relative meaning
along an abstract scale of utility, and modelers often compute WTP from
preference space coefficients to help make results more interpretable.

Perhaps less obvious is the fact that WTP estimates can be directly
compared with those from models estimated on other datasets since WTP
coefficients are independent of error scaling. This is particularly
convenient for comparing WTP estimates from different subsets of a
dataset. In a preference space model, parameters are proportional to
error scaling, and thus due to potential scale differences coefficients
estimated from different data sets cannot be directly compared
\citep{Swait2003, Helveston2018}. This also poses a challenge for
comparative studies or literature reviews that seek to compare outcomes
on similar topics across multiple studies.

Another perhaps less obvious implication of the preference space
parameterization is the assumption that distributions of marginal
utilities are independent across attributes, which induces a strong
correlation structure among WTP values \citep{Train2005}. This can make
it difficult to evaluate alternatives with different attribute levels
since WTP cannot be added across attributes. In a WTP space model, this
problem can be avoided by directly incorporating the correlation
structure among WTP coefficients, and as a result dollar values can be
summed to yield the total WTP for an alternative.

Finally, there is no theoretical basis for believing that marginal
utilities versus marginal WTPs should follow standard distributions
(e.g., normal and log-normal). In the absence of any theoretical basis
for these assumptions, the modeler is left to consider differences in
empirical outcomes, which as previously noted, there has not been much
definitive evidence that models in one space systematically out-perform
the other along all measures of significance.

\hypertarget{software-architecture-and-performance}{%
\section{Software architecture and
performance}\label{software-architecture-and-performance}}

\hypertarget{design-features-for-increased-estimation-speed}{%
\subsection{Design features for increased estimation
speed}\label{design-features-for-increased-estimation-speed}}

In maximum likelihood estimation (and simulated MLE for mixed logit
models), the log-likelihood function is computed many times as the
algorithm searches for parameters that minimize the negative of the
log-likelihood via gradient descent. The \pkg{logitr} package uses
several strategies to accelerate this process.

First, minimization of the negative log-likelihood is handled via the
\pkg{nloptr} package, which is an \proglang{R} interface to
\pkg{NLopt}---an open source program for nonlinear optimization started
by Steven G. Johnson \citep{Ypma2020}. One benefit of using \pkg{nloptr}
is that both the log-likelihood function and its gradient can be
computed within the same function. This reduces redundant computations
as many intermediate calculations are shared between the log-likelihood
and its gradient. Furthermore, analytic gradients are implemented for
both preference space and WTP space models and for multinomial and mixed
logit models.

Another important feature is that the choice probabilities are
reformulated to reduce the number of calculations needed to compute the
log-likelihood function. Instead of using Equation
\ref{eq:logitProbabilityMnl} to compute the probability of each
alternative in a choice set, the choice probability for the
\emph{chosen alternative}, \(P_{c}\), can be calculated as:

\begin{equation}
    P_{c} =
    \frac{
        1
    }{
        1 + \sum_{j \ne c}^J \exp(v_{j} - v_{c})
    },
    \label{eq:logitProbReformulated}
\end{equation}

\noindent  This results in a more stable and computationally faster
calculation of the log-likelihood, which is simplified from Equation
\ref{eq:logLikelihood} as

\begin{equation}
    L =
  \sum_n^N
  \sum_j^J
    \ln P_{nc}
    \label{eq:logLikelihoodReformulated}
\end{equation}

In addition, \pkg{logitr} takes advantage of the fact that, except for
the parameters, the data used in computing the log-likelihood function
and its gradient do not change, enabling a considerable amount of memory
reduction by pre-computing several intermediate computations that remain
constant throughout the estimation process. For example, the gradient
with respect to parameters \(\boldsymbol\uptheta\) of the log-likelihood
in equation \ref{eq:logLikelihoodReformulated} for multinomial logit
models can be written as follows:

\begin{equation}
  \begin{aligned}
    \frac{\partial \mathrm{L}}{\partial \boldsymbol\uptheta}
    = \sum_{n=1}^{N} - P_{nc} \left[\sum_{j \ne c}^J  \exp(v_{nj} - v_{nc}) \frac{\partial}{\partial \boldsymbol\uptheta} (v_{nj} - v_{nc})\right]\\
  \end{aligned}
    \label{eq:gradientReformulated}
\end{equation}

\noindent  In preference space models where
\(v_{nj} = \boldsymbol\upbeta^\top \mathrm{\mathbf{x}}_{nj} + \alpha p_{nj}\),
the partial derivative \(\partial / \partial \boldsymbol\uptheta\) in
equation \ref{eq:gradientReformulated} is:

\begin{equation}
    \frac{\partial}{\partial \alpha} (v_{nj} - v_{nc})
    = p_{nj} - p_{nc},
    \quad\quad
    \frac{\partial}{\partial \boldsymbol\upbeta} (v_{nj} - v_{nc})
    = \mathrm{\mathbf{x}}_{nj} - \mathrm{\mathbf{x}}_{nc}
    \label{eq:partialLPref}
\end{equation}

\noindent  In WTP space models where
\(v_{nj} = \lambda(\boldsymbol\upomega^\top \mathrm{\mathbf{x}}_{nj} - p_{nj})\),
the partial derivatives \(\partial / \partial \boldsymbol\uptheta\) in
equation \ref{eq:gradientReformulated} are:

\begin{equation}
    \frac{\partial}{\partial \lambda} (v_{nj} - v_{nc})
    = \frac{1}{\lambda}(v_{nj} - v_{nc}),
    \quad\quad
    \frac{\partial}{\partial \boldsymbol\omega} (v_{nj} - v_{nc})
    = \lambda (\mathrm{\mathbf{x}}_{nj} - \mathrm{\mathbf{x}}_{nc})
    \label{eq:partialLWtp}
\end{equation}

\noindent  The values of \(p_{nj} - p_{nc}\) and
\(\mathrm{\mathbf{x}}_{nj} - \mathrm{\mathbf{x}}_{nc}\) in Equations
\ref{eq:partialLPref} and \ref{eq:partialLWtp} are constant and can be
computed prior to starting the optimization loop. Furthermore, since
\(P_{nc}\), \((v_{nj} - v_{nc})\), and \(\exp(v_{nj} - v_{nc})\) are
already computed when calculating the log-likelihood, they can be used
to quickly compute the analytic gradient with only a few additional
calculations in each iteration of the algorithm.

Finally, the \texttt{parallel} package is also used to simultaneously
estimate multiple models from different starting points when estimating
a multi-start loop. For machines with multiple cores, this can
dramatically increase the size of the solution space searched without
substantially increasing estimation time.

\hypertarget{performance-benchmarking}{%
\subsection{Performance benchmarking}\label{performance-benchmarking}}

\begin{figure}[h]
\begin{center}
\includegraphics[width=0.8\linewidth]{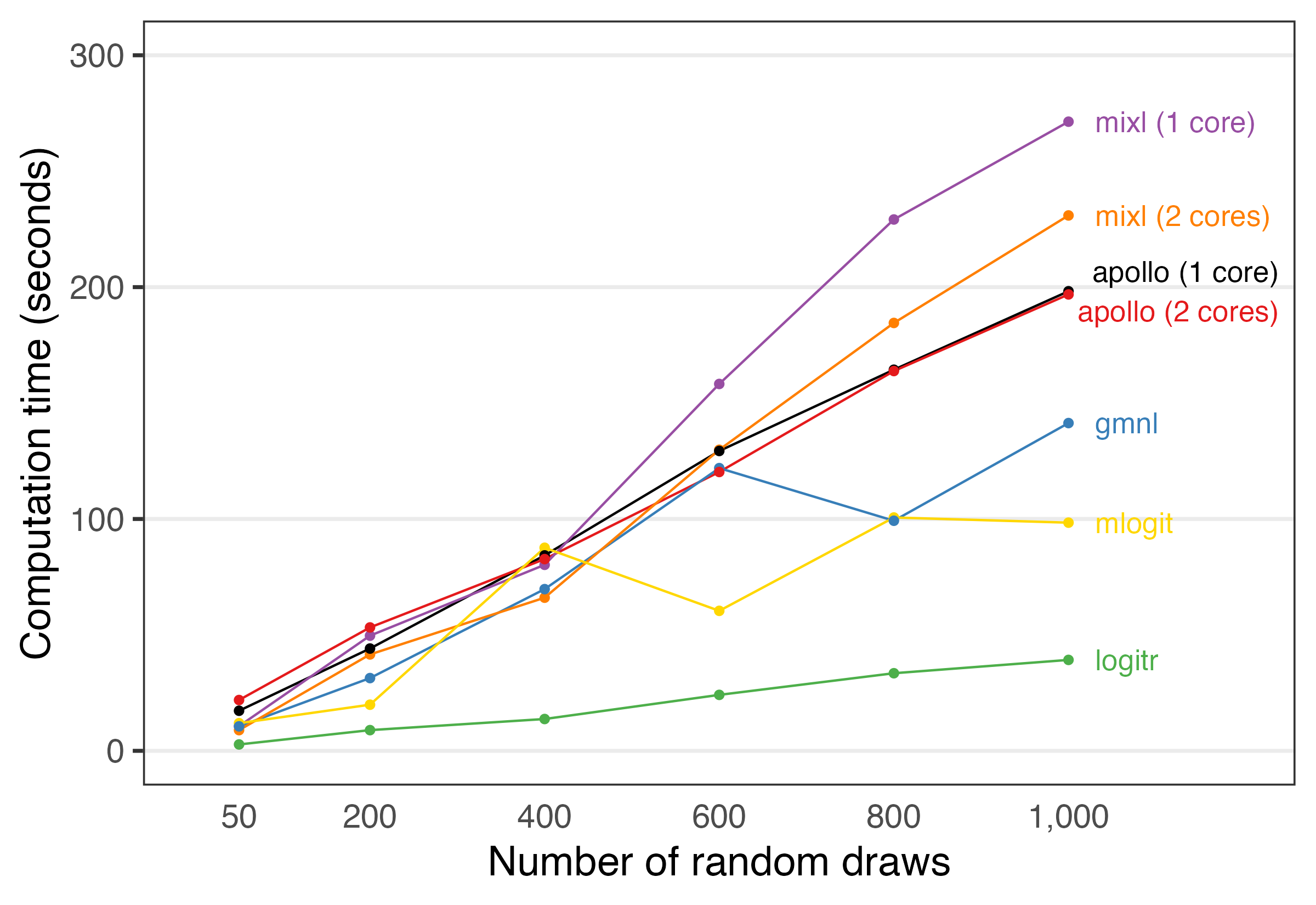}
\caption[Comparison of package estimation times for a preference space mixed logit model with four normally-distributed random parameters.]{Comparison of package estimation times for a preference space mixed logit model with four normally-distributed random parameters. The x-axis shows the number of random draws used in simulating the log-likelihood function.}
\label{fig:fig2}
\end{center}
\end{figure}

The design features implemented in \pkg{logitr} result in impressive
gains in overall efficiency compared to similar packages. To compare its
performance, a preference space mixed logit model was estimated using
\pkg{logitr}, \pkg{mlogit}, \pkg{mixl}, \pkg{gmnl}, and \pkg{apollo}.
Figure \ref{fig:fig2} shows the estimation time for each package plotted
against the number of random draws used in the mixed logit model. The
benchmark was carried out in a Google Colab notebook at
\url{https://colab.research.google.com/drive/1vYlBdJd4xCV43UwJ33XXpO3Ys8xWkuxx?usp=sharing}.

With just 50 random draws, \pkg{logitr} is particularly fast, clocking
in at 4.3 times faster than \pkg{mixl} with one core, 3.6 times faster
than \pkg{mixl} with two cores, 5.1 times faster than \pkg{mlogit}, 4.5
times faster than \pkg{gmnl}, 8.2 times faster than \pkg{apollo} with
one core, and 10.6 times faster than \pkg{apollo} with two cores. The
comparative difference decreases with higher numbers of draws, though
even at 1,000 draws \pkg{logitr} is still over twice as fast as
\pkg{mlogit}, the next-fastest package in the benchmark (see Table 1).
Speeds could potentially be further improved by parallelizing elements
of the gradient calculation. In addition, it is important to note that
\pkg{mixl} in particular may perform better at very large numbers of
draws (e.g.~10,000 or more) if used on a machine with a large number of
cores.

\newpage

\begin{center}
Table 1: Comparison of package estimation times for a preference space mixed logit model with four normally-distributed random parameters.
\end{center}

\begin{CodeChunk}
\captionsetup[table]{labelformat=empty,skip=1pt}
\begin{longtable}{lrrrrrrrrrrrr}
\toprule
 & \multicolumn{6}{c}{Estimation time (s)} & \multicolumn{6}{c}{Times slower than logitr} \\ 
 \cmidrule(lr){2-7} \cmidrule(lr){8-13}
   & 50 & 200 & 400 & 600 & 800 & 1000 &  50 &  200 &  400 &  600 &  800 &  1000 \\ 
\midrule
logitr & 3 & 9 & 14 & 24 & 33 & 39 & 1.0 & 1.0 & 1.0 & 1.0 & 1.0 & 1.0 \\ 
mixl (1 core) & 11 & 50 & 80 & 158 & 229 & 271 & 3.9 & 5.6 & 5.8 & 6.6 & 6.8 & 6.9 \\ 
mixl (2 cores) & 9 & 42 & 66 & 130 & 185 & 231 & 3.2 & 4.7 & 4.8 & 5.4 & 5.5 & 5.9 \\ 
mlogit & 12 & 20 & 88 & 60 & 101 & 98 & 4.3 & 2.2 & 6.4 & 2.5 & 3.0 & 2.5 \\ 
gmnl & 11 & 31 & 70 & 122 & 99 & 141 & 3.8 & 3.5 & 5.1 & 5.1 & 3.0 & 3.6 \\ 
apollo (1 core) & 17 & 44 & 84 & 129 & 164 & 198 & 6.3 & 4.9 & 6.1 & 5.4 & 4.9 & 5.1 \\ 
apollo (2 cores) & 22 & 53 & 83 & 120 & 164 & 197 & 8.0 & 6.0 & 6.0 & 5.0 & 4.9 & 5.0 \\ 
 \bottomrule
\end{longtable}
\end{CodeChunk}

\newpage

\section[logitr package]{Using the \pkg{logitr}
package}\label{logitr-package}

\hypertarget{installation}{%
\subsection{Installation}\label{installation}}

The \pkg{logitr} package can be installed from CRAN:

\begin{CodeChunk}
\begin{CodeInput}
R> install.packages("logitr")
\end{CodeInput}
\end{CodeChunk}

\noindent The development version can be installed from GitHub using the
\pkg{remotes} package:

\begin{CodeChunk}
\begin{CodeInput}
R> remotes::install_github("jhelvy/logitr")
\end{CodeInput}
\end{CodeChunk}

\noindent The package is loaded in \proglang{R} with:

\begin{CodeChunk}
\begin{CodeInput}
R> library("logitr")
\end{CodeInput}
\end{CodeChunk}

\hypertarget{data-format}{%
\subsection{Data format}\label{data-format}}

The \pkg{logitr} package requires that data be structured in a
\texttt{data.frame} and arranged in a ``long'' format
\citep{Wickham2014} where each row contains data on a single alternative
from a choice observation. The choice observations do not have to be
symmetric, meaning they can have a ``ragged'' structure where different
choice observations have different numbers of alternatives. The data
must include variables for each of the following:

\begin{itemize}
\tightlist
\item
  Outcome: a dummy-coded variable that identifies which alternative was
  chosen (\texttt{1} is chosen, \texttt{0} is not chosen). Only one
  alternative should have a \texttt{1} per choice observation.
\item
  Observation ID: a sequence of repeated numbers that identifies each
  unique choice observation. For example, if the first three choice
  observations had 2 alternatives each, then the first 6 rows of the
  \texttt{obsID} variable would be \texttt{1,\ 1,\ 2,\ 2,\ 3,\ 3}.
\item
  Covariates: other variables that will be used as model covariates.
\end{itemize}

\noindent The \pkg{logitr} package contains several example data sets
that illustrate this structure. The \texttt{yogurt} data set will be
used as a running example throughout this paper to illustrate the key
features of the package. The data set contains observations of yogurt
purchases by a panel of 100 households \citep{Jain1994}. Choice is
identified by the \texttt{choice} column, each choice observation is
identified by the \texttt{obsID} column, and the columns \texttt{price},
\texttt{feat}, and \texttt{brand} can be used as model covariates:

\begin{CodeChunk}
\begin{CodeInput}
R> head(yogurt)
\end{CodeInput}
\begin{CodeOutput}
# A tibble: 6 x 7
     id obsID   alt choice price  feat brand  
  <dbl> <int> <int>  <dbl> <dbl> <dbl> <chr>  
1     1     1     1      0  8.1      0 dannon 
2     1     1     2      0  6.10     0 hiland 
3     1     1     3      1  7.90     0 weight 
4     1     1     4      0 10.8      0 yoplait
5     1     2     1      1  9.80     0 dannon 
6     1     2     2      0  6.40     0 hiland 
\end{CodeOutput}
\end{CodeChunk}

This data set also includes an \texttt{alt} variable that determines the
alternatives included in the choice set of each observation and an
\texttt{id} variable that determines the individual as the data have a
panel structure containing multiple choice observations from each
individual.

\hypertarget{model-specification-interface}{%
\subsection{Model specification
interface}\label{model-specification-interface}}

Models are specified and estimated using the \texttt{logitr()} function.
The \texttt{data} argument should be set to the data frame containing
the data, and the \texttt{outcome} and \texttt{obsID} arguments should
be set to the column names in the data frame that correspond to the
dummy-coded outcome (choice) variable and the observation ID variable,
respectively. All variables to be used as model covariates should be
provided as a vector of column names to the \texttt{pars} argument. Each
variable in the vector is additively included as a covariate in the
utility model, with the interpretation that they represent utilities in
preference space models and WTPs in a WTP space model.

For example, consider a preference space model where the utility for
yogurt is given by the following utility model:

\begin{equation}
    u_{j} =
        \alpha p_{j} +
        \beta_1 x_{j1} +
        \beta_2 x_{j2} +
        \beta_3 x_{j3} +
        \beta_4 x_{j4} +
        \varepsilon_{j},
\label{eq:yogurtUtilityPref}
\end{equation}

\noindent where \(p_{j}\) is \texttt{price}, \(x_{j1}\) is
\texttt{feat}, and \(x_{j2-4}\) are dummy-coded variables for each
\texttt{brand} (with the fourth brand representing the reference level).
This model can be estimated using the \texttt{logitr()} function as
follows:

\begin{CodeChunk}
\begin{CodeInput}
R> mnl_pref <- logitr(
+   data    = yogurt,
+   outcome = "choice",
+   obsID   = "obsID",
+   pars    = c("price", "feat", "brand")
+ )
\end{CodeInput}
\end{CodeChunk}

The equivalent model in the WTP space is given by the following utility
model:

\begin{equation}
    u_{j} =
        \lambda \left(
            \omega_1 x_{j1} +
            \omega_1 x_{j2} +
            \omega_1 x_{j3} +
            \omega_2 x_{j4} -
            p_{j}
        \right) +
        \varepsilon_{j},
\label{eq:yogurtUtilityWtp}
\end{equation}

\noindent To specify this model, simply move \texttt{"price"} from the
\texttt{pars} argument to the \texttt{scalePar} argument:

\begin{CodeChunk}
\begin{CodeInput}
R> mnl_wtp <- logitr(
+   data     = yogurt,
+   outcome  = "choice",
+   obsID    = "obsID",
+   pars     = c("feat", "brand"),
+   scalePar = "price"
+ )
\end{CodeInput}
\end{CodeChunk}

\noindent  In the above model, the variables in \texttt{pars} are
marginal WTPs, whereas in the preference space model they are marginal
utilities. Price is separately specified with the \texttt{scalePar}
argument because it acts as a scaling term in WTP space models. While
price is the most typical scaling variable, other continuous variables
can also be used; for example, time could be used to obtain marginal
estimates of ``willingness to wait.''

Interactions between covariates can be entered in the \texttt{pars}
vector separated by the \texttt{*} symbol. For example, an interaction
between \texttt{price} with \texttt{feat} in the above preference space
model could be included by specifying
\texttt{pars\ =\ c("price",\ "feat",\ "brand",\ "price*feat")}, or even
more concisely just \texttt{pars\ =\ c("price*feat",\ "brand")} as the
interaction between \texttt{price} and \texttt{feat} will produce
individual parameters for \texttt{price} and \texttt{feat} in addition
to the interaction parameter.

Although the \pkg{logitr} model specification interface is a departure
from the popular \texttt{formula} interface used in other similar
packages such as \pkg{mlogit}, it was designed to be more uniform and
streamlined for estimating either preference or WTP space models. For
WTP space models in particular, using the \texttt{formula} interface can
be confusing as it requires that covariates be additively specified
(e.g., \texttt{choice\ \textasciitilde{}\ price\ +\ feat\ +\ brand}),
which is inconsistent with the underlying WTP space utility model
parameterization in which the price parameter (\(\lambda\)) scales the
WTP parameters.

For example, consider how the \texttt{formula} interface is used in the
\pkg{gmnl} package. Using \pkg{gmnl}, WTP space models are specified by
1) modifying the price attribute in the data to be the negative of price
prior to estimation, 2) specifying the \texttt{model} argument as
\texttt{"smnl"}, 3) additively including all parameters in the
\texttt{formula} (including price) along with appropriate \texttt{0}s
and \texttt{1}s in the additional formula components to properly specify
a scaled multinomial logit model, 4) specifying a vector of
\texttt{TRUE} and \texttt{FALSE} values to the \texttt{fixed} argument
for every model parameter, and 5) providing starting values where the
price / scale parameter is set to \texttt{1} for stability (otherwise
the default value of \texttt{0} will be used, which usually results in a
convergence failure). To estimate the previous example WTP space model
using \pkg{gmnl}, the data would first need to be formatted using the
\texttt{mlogit.data()} function:

\begin{CodeChunk}
\begin{CodeInput}
R> data_gmnl <- mlogit.data(
+     data     = yogurt,
+     shape    = "long",
+     choice   = "choice",
+     id.var   = 'id',
+     alt.var  = 'alt',
+     chid.var = 'obsID',
+     opposite = 'price'
+ )
\end{CodeInput}
\end{CodeChunk}

The model would then be estimated using the \texttt{gmnl()} function:

\begin{CodeChunk}
\begin{CodeInput}
R> mnl_wtp <- gmnl(
+   data = data_gmnl,
+   formula = choice ~ price + feat + brand | 0 | 0 | 0 | 1,
+   fixed = c(TRUE, FALSE, FALSE, FALSE, FALSE, TRUE, FALSE),
+   model = "smnl", method = "bhhh",
+   start = c(1, 0, 0, 0, 0, 0, 0)
+ )
\end{CodeInput}
\end{CodeChunk}

Compared to the \pkg{logitr} interface, the above syntax is considerably
more complex, and it is also not obvious that this specification will
produce WTP estimates. The additive inclusion of \texttt{price} in
\texttt{formula} is particularly confusing for a model that produces WTP
estimates as it is inconsistent with the WTP space utility model.
Finally, if the user fails to remember the specific set of preparation
steps prior to estimation (such as taking the negative of price),
results could be confusing.

The \pkg{apollo} and \pkg{mixl} packages can also be used to estimate
WTP space models, but they require that the user hand-specify the
utility model either as a function or string. While this provides
greater flexibility in the types of models that can be estimated, it
also requires much more effort by the user to carefully specify every
model. Simple modifications, such as adding in one more variable,
require that the user modify multiple settings, including modifying the
starting parameter vector as well as re-defining the utility model
function or string (among other potential required changes). Even with
helpful guides and examples provided by the developers of these
packages, more effort is required by the user to appropriately use these
packages to estimate even the simplest of models.

These issues motivated the use of an alternative model specification
interface for package \pkg{logitr}, with the goal of developing a syntax
that is at least as intuitive as the \texttt{formula} interface but more
uniform for estimating models in either the preference or WTP space.

\hypertarget{continuous-and-discrete-variable-coding}{%
\subsection{Continuous and discrete variable
coding}\label{continuous-and-discrete-variable-coding}}

Variables are modeled in \pkg{logitr} as either continuous or discrete
based on their data type. Numeric variables are modeled with a single
``slope'' coefficient, and \texttt{character} or \texttt{factor} type
variables are modeled as categorical variables with dummy-coded
coefficients for all but the first level, which serves as the reference
level. For example, since the \texttt{price} variable in the
\texttt{yogurt} data frame is a numeric variable, it will be modeled
with a single coefficient representing the change in utility for
marginal changes in price. In contrast, since \texttt{brand} is a
character type with the levels \texttt{"dannon"}, \texttt{"hiland"},
\texttt{"weight"}, and \texttt{"yoplait"}, it will be modeled with three
dummy-coded coefficients with the \texttt{"dannon"} brand set as the
reference level as it is alphabetically first.

To change the reference level for discrete variables, modify the factor
levels for that variable prior to model estimation with the
\texttt{factor()} function. For example, the following code will set
\texttt{"weight"} instead of \texttt{"dannon"} as the reference level
for the \texttt{brand} variable:

\begin{CodeChunk}
\begin{CodeInput}
R> yogurt2 <- logitr::yogurt
R> yogurt2$brand <- factor(
+     x = yogurt2$brand, 
+     levels = c("weight", "hiland", "yoplait", "dannon")
+ )
R> levels(yogurt2$brand)
\end{CodeInput}
\begin{CodeOutput}
[1] "weight"  "hiland"  "yoplait" "dannon" 
\end{CodeOutput}
\end{CodeChunk}

Any variable can be made discrete by either converting it to a
\texttt{character} or \texttt{factor} type prior to model estimation or
by creating new dummy-coded variables for each level and including all
but the reference level as model covariates. A recommended approach is
to use the \texttt{dummy\_cols()} function from the \pkg{fastDummies}
package, which generates dummy-coded variables for each unique level of
a discrete variable \citep{Kaplan2020}.

\hypertarget{estimating-multinomial-logit-models}{%
\subsection{Estimating multinomial logit
models}\label{estimating-multinomial-logit-models}}

The \texttt{logitr()} function estimates preference space models by
default. Once a model is estimated, the \texttt{summary()} function can
be used to print a summary of the estimated model results to the
console. For example, consider again the preference space model with
\texttt{price}, \texttt{feat}, and \texttt{brand} as model covariates:

\begin{CodeChunk}
\begin{CodeInput}
R> mnl_pref <- logitr(
+   data    = yogurt,
+   outcome = "choice",
+   obsID   = "obsID",
+   pars    = c("price", "feat", "brand")
+ )
\end{CodeInput}
\end{CodeChunk}

\begin{CodeChunk}
\begin{CodeInput}
R> summary(mnl_pref)
\end{CodeInput}
\begin{CodeOutput}
=================================================

Model estimated on: Tue Oct 04 11:58:55 2022 

Using logitr version: 0.8.0 

Call:
logitr(data = yogurt, outcome = "choice", obsID = "obsID", pars = c("price", 
    "feat", "brand"))

Frequencies of alternatives:
       1        2        3        4 
0.402156 0.029436 0.229270 0.339138 

Exit Status: 3, Optimization stopped because ftol_rel or ftol_abs was reached.
                                
Model Type:    Multinomial Logit
Model Space:          Preference
Model Run:                1 of 1
Iterations:                   21
Elapsed Time:        0h:0m:0.02s
Algorithm:        NLOPT_LD_LBFGS
Weights Used?:             FALSE
Robust?                    FALSE

Model Coefficients: 
              Estimate Std. Error  z-value  Pr(>|z|)    
price        -0.366555   0.024365 -15.0441 < 2.2e-16 ***
feat          0.491439   0.120062   4.0932 4.254e-05 ***
brandhiland  -3.715477   0.145417 -25.5506 < 2.2e-16 ***
brandweight  -0.641138   0.054498 -11.7645 < 2.2e-16 ***
brandyoplait  0.734519   0.080642   9.1084 < 2.2e-16 ***
---
Signif. codes:  0 '***' 0.001 '**' 0.01 '*' 0.05 '.' 0.1 ' ' 1
                                     
Log-Likelihood:         -2656.8878790
Null Log-Likelihood:    -3343.7419990
AIC:                     5323.7757580
BIC:                     5352.7168000
McFadden R2:                0.2054148
Adj McFadden R2:            0.2039195
Number of Observations:  2412.0000000
\end{CodeOutput}
\end{CodeChunk}

The summary includes information about the \texttt{logitr()} function
call, the frequency of chosen alternatives, the optimization exit
status, the estimated coefficients, the log-likelihood value at the
solution, and several measures of model fit. In this example, three
coefficients were estimated for the ``brand'' attribute, with
\texttt{"dannon"} set as the default reference level.

These results indicate that, all else being equal, people in this sample
on average preferred the ``Yoplait'' brand the most, followed by
``Dannon'' (the reference level, which would be 0 on the utility scale),
``Weight Watchers'', and finally ``Hiland''. The results also indicate
that utility decreased by a value of -0.367 for every dollar increase in
price, which is logically consistent with people preferring lower rather
than higher prices, all else being equal.

Values from an estimated model can be extracted using methods designed
for objects of class \texttt{logitr}, such as the following:

The estimated coefficients:

\begin{CodeChunk}
\begin{CodeInput}
R> coef(mnl_pref)
\end{CodeInput}
\begin{CodeOutput}
       price         feat  brandhiland  brandweight brandyoplait 
  -0.3665546    0.4914392   -3.7154773   -0.6411384    0.7345195 
\end{CodeOutput}
\end{CodeChunk}

The coefficient standard errors:

\begin{CodeChunk}
\begin{CodeInput}
R> se(mnl_pref)
\end{CodeInput}
\begin{CodeOutput}
       price         feat  brandhiland  brandweight brandyoplait 
  0.02436526   0.12006175   0.14541671   0.05449794   0.08064229 
\end{CodeOutput}
\end{CodeChunk}

The log-likelihood:

\begin{CodeChunk}
\begin{CodeInput}
R> logLik(mnl_pref)
\end{CodeInput}
\begin{CodeOutput}
'log Lik.' -2656.888 (df=5)
\end{CodeOutput}
\end{CodeChunk}

The variance-covariance matrix:

\begin{CodeChunk}
\begin{CodeInput}
R> vcov(mnl_pref)
\end{CodeInput}
\begin{CodeOutput}
                     price          feat  brandhiland  brandweight
price         0.0005936657  5.729619e-04  0.001851795 1.249988e-04
feat          0.0005729619  1.441482e-02  0.000855011 5.092011e-06
brandhiland   0.0018517954  8.550110e-04  0.021146019 1.490080e-03
brandweight   0.0001249988  5.092011e-06  0.001490080 2.970026e-03
brandyoplait -0.0015377721 -1.821331e-03 -0.003681036 7.779428e-04
              brandyoplait
price        -0.0015377721
feat         -0.0018213311
brandhiland  -0.0036810363
brandweight   0.0007779427
brandyoplait  0.0065031782
\end{CodeOutput}
\end{CodeChunk}

\hypertarget{estimating-willingness-to-pay}{%
\subsection{Estimating willingness to
pay}\label{estimating-willingness-to-pay}}

Coefficients in preference space models reflect marginal changes in
utility, which only have relative value. To make these coefficients more
interpretable, modelers often divide the non-price parameters by the
negative of the price parameter to obtain estimates of WTP (the negative
of the price parameter is used so that marginal WTPs have a positive
interpretation). This can be computed using the \texttt{wtp()} function:

\begin{CodeChunk}
\begin{CodeInput}
R> wtp(mnl_pref, scalePar = "price")
\end{CodeInput}
\begin{CodeOutput}
               Estimate Std. Error  z-value  Pr(>|z|)    
scalePar       0.366555   0.024378  15.0361 < 2.2e-16 ***
feat           1.340699   0.360539   3.7186 0.0002003 ***
brandhiland  -10.136219   0.583206 -17.3802 < 2.2e-16 ***
brandweight   -1.749094   0.181960  -9.6125 < 2.2e-16 ***
brandyoplait   2.003848   0.143323  13.9813 < 2.2e-16 ***
---
Signif. codes:  0 '***' 0.001 '**' 0.01 '*' 0.05 '.' 0.1 ' ' 1
\end{CodeOutput}
\end{CodeChunk}

The \texttt{wtp()} function returns a data frame of the WTP estimates
and standard errors. The coefficient labeled \texttt{scalePar} is the
negative of the price coefficient from the preference space model, which
can also be interpreted as the scale coefficient in the WTP space model
as it scales all the WTP coefficients. Standard errors are estimated
using the Krinsky and Robb parametric bootstrapping method
\citep{Krinsky1986}.

In contrast to the preference space model coefficients, the WTP
estimates above have units of currency (in this case \$US dollars) and
can be interpreted as how much the average person in the sample would be
willing to pay for each feature, all else being equal. For example, the
brand coefficients suggest that, relative to the ``Dannon'' brand,
consumers are on average willing to pay an additional \$2.00 for the
``Yoplait'' brand, -\$1.75 for the ``Weight Watchers'' brand, and
-\$10.14 for the ``Hiland'' brand (negative WTPs indicate a relative
preference for ``Dannon'').

WTPs can also be directly estimated using a WTP space model. In this
case, the \texttt{pars} argument should contain only attributes for
which WTPs are to be estimate. The variable for ``price'' should be
provided separately using the \texttt{scalePar} argument. For example,
consider again the WTP space model with \texttt{feat} and \texttt{brand}
WTP covariates:

\newpage

\begin{CodeChunk}
\begin{CodeInput}
R> set.seed(123)
R> 
R> mnl_wtp <- logitr(
+   data     = yogurt,
+   outcome  = "choice",
+   obsID    = "obsID",
+   pars     = c("feat", "brand"),
+   scalePar = "price",
+   numMultiStarts = 10,
+   numCores = 1
+ )
\end{CodeInput}
\end{CodeChunk}

\noindent In the above example, a 10-iteration multi-start optimization
loop was implemented by setting \texttt{numMultiStarts\ =\ 10}. This
runs the minimization of the negative log-likelihood function 10 times
from 10 different sets of random starting points (the first iteration
uses all \texttt{0}s except for the price parameter which starts at
\texttt{1}). This is recommended as WTP space models have a non-convex
log-likelihood function and thus could have multiple local minimia. Note
also that the multi-start loop can be parallelized by setting
\texttt{numCores} to an integer greater than 1. The default value is one
less than the total number of available cores, but
\texttt{numCores\ =\ 1} is used here to ensure reproducibility.

The WTP estimates have the same interpretation as those computed from
the preference space model. In the summary output, a short summary of
each iteration of the multi-start loop is provided first followed by the
same summary information about the preference space model. Because a
multi-start loop was used, only the summary of the ``best'' estimated
model is returned (determined by the iteration with the largest
log-likelihood value):

\begin{CodeChunk}
\begin{CodeInput}
R> summary(mnl_wtp)
\end{CodeInput}
\begin{CodeOutput}
=================================================

Model estimated on: Tue Oct 04 11:58:55 2022 

Using logitr version: 0.8.0 

Call:
logitr(data = yogurt, outcome = "choice", obsID = "obsID", pars = c("feat", 
    "brand"), scalePar = "price", numMultiStarts = 10, numCores = 1)

Frequencies of alternatives:
       1        2        3        4 
0.402156 0.029436 0.229270 0.339138 

Summary Of Multistart Runs:
   Log Likelihood Iterations Exit Status
1       -2656.888         38           3
2       -2656.888         52           3
3       -2656.888         59           3
4       -2656.888         40           3
5       -2656.888         44           3
6       -2656.888         44           3
7       -2656.888         63           3
8       -2656.888         36           3
9       -2656.888         40           3
10      -2656.888         45           3

Use statusCodes() to view the meaning of each status code

Exit Status: 3, Optimization stopped because ftol_rel or ftol_abs was reached.
                                 
Model Type:     Multinomial Logit
Model Space:   Willingness-to-Pay
Model Run:                1 of 10
Iterations:                    38
Elapsed Time:         0h:0m:0.04s
Algorithm:         NLOPT_LD_LBFGS
Weights Used?:              FALSE
Robust?                     FALSE

Model Coefficients: 
               Estimate Std. Error  z-value  Pr(>|z|)    
scalePar       0.366583   0.024366  15.0448 < 2.2e-16 ***
feat           1.340593   0.355867   3.7671 0.0001651 ***
brandhiland  -10.135764   0.576089 -17.5941 < 2.2e-16 ***
brandweight   -1.749083   0.179898  -9.7226 < 2.2e-16 ***
brandyoplait   2.003821   0.142377  14.0740 < 2.2e-16 ***
---
Signif. codes:  0 '***' 0.001 '**' 0.01 '*' 0.05 '.' 0.1 ' ' 1
                                     
Log-Likelihood:         -2656.8878779
Null Log-Likelihood:    -3343.7419990
AIC:                     5323.7757559
BIC:                     5352.7168000
McFadden R2:                0.2054148
Adj McFadden R2:            0.2039195
Number of Observations:  2412.0000000
\end{CodeOutput}
\end{CodeChunk}

In the above summary, iteration 1 converged to a solution with a
log-likelihood value of -2656.888. While all of the iterations arrived
at the same solution in this particular example, this is not always the
case nor is it guaranteed. Because the previous examples are both fixed
parameter models, the WTP estimates from the WTP space model are
identical to those computed from the preference space model. The WTPs
from each model can be quickly compared using the \texttt{wtpCompare()}
function:

\newpage

\begin{CodeChunk}
\begin{CodeInput}
R> wtpCompare(
+   model_pref = mnl_pref,
+   model_wtp  = mnl_wtp,
+   scalePar   = "price"
+ )
\end{CodeInput}
\begin{CodeOutput}
                      pref           wtp  difference
scalePar         0.3665546     0.3665832  0.00002867
feat             1.3406987     1.3405926 -0.00010605
brandhiland    -10.1362190   -10.1357635  0.00045548
brandweight     -1.7490940    -1.7490826  0.00001133
brandyoplait     2.0038476     2.0038208 -0.00002686
logLik       -2656.8878790 -2656.8878779  0.00000106
\end{CodeOutput}
\end{CodeChunk}

In the above summary, the \texttt{pref} column contains the computed
WTPs from the preference space model, and the \texttt{wtp} column
contains the directly estimated WTPs from the WTP space model. The
\texttt{difference} column is the computed difference between them. This
is a helpful tool for assessing whether the two models converged to the
same solution.

\hypertarget{estimating-mixed-logit-models}{%
\subsection{Estimating mixed logit
models}\label{estimating-mixed-logit-models}}

The mixed logit model is a popular approach for modeling unobserved
heterogeneity across individuals, which is implemented by assuming that
parameters vary randomly across individuals according to a chosen
distribution \citep{McFadden2000}. A mixed logit model is specified by
setting the \texttt{randPars} argument in the \texttt{logitr()} function
equal to a named vector defining parameter distributions. The current
package version (0.8.0) supports the following distributions:

\begin{itemize}
\tightlist
\item
  Normal: \texttt{"n"}
\item
  Log-normal: \texttt{"ln"}
\item
  Zero-censored normal: \texttt{"cn"}
\end{itemize}

Mixed logit models will estimate a mean and standard deviation of the
underlying normal distribution for each random coefficient. Note that
log-normal or zero-censored normal parameters force positivity, so when
using these it is often necessary to use the negative of a value
(e.g.~for ``price'', which typically has a negative coefficient). Mixed
logit models in \pkg{logitr} assume uncorrelated heterogeneity
covariances by default, though full covariances can be estimated using
the \texttt{correlation\ =\ TRUE} argument. For WTP space models, the
\texttt{scalePar} parameter can also be modeled as following a random
distribution by setting the \texttt{randScale} argument equal to
\texttt{"n"}, \texttt{"ln"}, or \texttt{"cn"}.

The code below is an example of a mixed logit model where the observed
utility for yogurts is
\(v_{j} = \alpha p_{j} + \beta_1 x_{j1} + \beta_2 x_{j2} + \beta_3 x_{j3} + \beta_4 x_{j4}\),
where \(p_{j}\) is \texttt{price}, \(x_{j1}\) is \texttt{feat}, and
\(x_{j2-4}\) are dummy-coded variables for \texttt{brand}. To model
\texttt{feat} as well as each of the brands as normally-distributed, set
\texttt{randPars\ =\ c(feat\ =\ "n",\ brand\ =\ "n")}. Since mixed logit
models have a non-convex log-likelihood function, it is recommended to
use a multi-start search to run the optimization multiple times from
different random starting points. Mixed logit models typically take
longer to estimate than fixed parameter models, so setting a larger
number for \texttt{numMultiStarts} could take several minutes to
complete.

Note that since the \texttt{yogurt} data has a panel structure
(i.e.~multiple choice observations for each respondent), it is necessary
to set the \texttt{panelID} argument to the \texttt{id} variable, which
identifies the individual. This will use the panel version of the
log-likelihood (see \citep{Train2009} section 6.7 for details).

\begin{CodeChunk}
\begin{CodeInput}
R> set.seed(456)
R> 
R> mxl_pref <- logitr(
+   data     = yogurt,
+   outcome  = 'choice',
+   obsID    = 'obsID',
+   panelID  = 'id',
+   pars     = c('price', 'feat', 'brand'),
+   randPars = c(feat = 'n', brand = 'n'),
+   numMultiStarts = 10,
+   numCores = 1
+ )
\end{CodeInput}
\end{CodeChunk}

\begin{CodeChunk}
\begin{CodeInput}
R> summary(mxl_pref)
\end{CodeInput}
\begin{CodeOutput}
=================================================

Model estimated on: Tue Oct 04 11:58:57 2022 

Using logitr version: 0.8.0 

Call:
logitr(data = yogurt, outcome = "choice", obsID = "obsID", pars = c("price", 
    "feat", "brand"), randPars = c(feat = "n", brand = "n"), 
    panelID = "id", numMultiStarts = 10, numCores = 1)

Frequencies of alternatives:
       1        2        3        4 
0.402156 0.029436 0.229270 0.339138 

Summary Of Multistart Runs:
   Log Likelihood Iterations Exit Status
1       -1266.550         34           3
2       -1300.751         64           3
3       -1260.216         35           3
4       -1261.216         43           3
5       -1269.066         40           3
6       -1239.294         56           3
7       -1343.221         59           3
8       -1260.006         55           3
9       -1273.143         52           3
10      -1304.384         59           3

Use statusCodes() to view the meaning of each status code

Exit Status: 3, Optimization stopped because ftol_rel or ftol_abs was reached.
                             
Model Type:       Mixed Logit
Model Space:       Preference
Model Run:            6 of 10
Iterations:                56
Elapsed Time:        0h:0m:2s
Algorithm:     NLOPT_LD_LBFGS
Weights Used?:          FALSE
Panel ID:                  id
Robust?                 FALSE

Model Coefficients: 
                 Estimate Std. Error  z-value  Pr(>|z|)    
price           -0.448338   0.039987 -11.2120 < 2.2e-16 ***
feat             0.776990   0.193521   4.0150 5.944e-05 ***
brandhiland     -6.367360   0.520828 -12.2255 < 2.2e-16 ***
brandweight     -3.668683   0.307207 -11.9421 < 2.2e-16 ***
brandyoplait     1.122492   0.203483   5.5164 3.460e-08 ***
sd_feat          0.567495   0.225004   2.5222   0.01166 *  
sd_brandhiland  -3.181844   0.371697  -8.5603 < 2.2e-16 ***
sd_brandweight   4.097130   0.232495  17.6225 < 2.2e-16 ***
sd_brandyoplait  3.261281   0.219902  14.8306 < 2.2e-16 ***
---
Signif. codes:  0 '***' 0.001 '**' 0.01 '*' 0.05 '.' 0.1 ' ' 1
                                     
Log-Likelihood:         -1239.2944250
Null Log-Likelihood:    -3343.7419990
AIC:                     2496.5888500
BIC:                     2548.6828000
McFadden R2:                0.6293690
Adj McFadden R2:            0.6266774
Number of Observations:  2412.0000000

Summary of 10k Draws for Random Coefficients: 
             Min.    1st Qu.     Median       Mean    3rd Qu. Max.
feat         -Inf  0.3938347  0.7765564  0.7761956  1.1591475  Inf
brandhiland  -Inf -8.5118797 -6.3663393 -6.3644102 -4.2201174  Inf
brandweight  -Inf -6.4342648 -3.6720435 -3.6750045 -0.9090452  Inf
brandyoplait -Inf -1.0817673  1.1169084  1.1141118  3.3162383  Inf
\end{CodeOutput}
\end{CodeChunk}

Since the \texttt{feat} and \texttt{brand} attributes were modeled as
normally distributed across the population, each of these covariates
have two parameters that describe the mean and standard deviation of a
normal distribution. For example, the coefficients \texttt{brandyoplait}
and \texttt{sd\_brandyoplait} indicate that, relative to the ``Dannon''
brand, the marginal utility for the ``Yoplait'' brand follows a normal
distribution across the population where
\(\beta_1 \sim \mathcal{N} (\mu = 1.122, \sigma = 3.261)\). For mixed
logit models, a summary of all random parameter distributions is printed
at the bottom of the summary output. In this example, there appears to
be less heterogeneity in preferences for the ``Yoplait'' and ``Hiland''
brands compared to the ``Weight Watchers'' brand, which has a larger
standard deviation parameter.

Note that mixed logit models can sometimes produce negative values for
the standard deviation parameters; in these cases, the parameters should
be interpreted as positive. The negative values are an artifact of how
the simulated MLE algorithm works. Since the normal distribution is
symmetric, it does not matter if draws are generated with
\(\mu + \sigma Z\) or \(\mu - \sigma Z\), where \(Z\) is a standard
normal and \(\mu\) and \(\sigma\) are the mean and standard deviation
parameters. By allowing the standard deviation parameter to be negative,
the optimization is unconstrained, making it a much easier problem to
solve.

\hypertarget{estimating-wtp-space-mixed-logit-models}{%
\subsection{Estimating WTP space mixed logit
models}\label{estimating-wtp-space-mixed-logit-models}}

WTP space mixed logit models have the advantage of being able to
directly specify the assumed distribution of WTP across the population.
As with fixed parameter models, estimating a mixed logit WTP space model
requires that price be separately specified with the \texttt{scalePar}
argument. The \texttt{randPars} argument is used to specify random
parameters. For example, the following can be used to estimate a model
where the WTP distributions for \texttt{feat} and each \texttt{brand} is
assumed to be normally distributed:

\begin{CodeChunk}
\begin{CodeInput}
R> set.seed(6789)
R> 
R> mxl_wtp <- logitr(
+   data     = yogurt,
+   outcome  = "choice",
+   obsID    = "obsID",
+   panelID  = 'id',
+   pars     = c("feat", "brand"),
+   scalePar = "price",
+   randPars = c(feat = "n", brand = "n"),
+   numMultiStarts = 10,
+   numCores = 1
+ )
\end{CodeInput}
\end{CodeChunk}

\begin{CodeChunk}
\begin{CodeInput}
R> summary(mxl_wtp)
\end{CodeInput}
\begin{CodeOutput}
=================================================

Model estimated on: Tue Oct 04 11:59:16 2022 

Using logitr version: 0.8.0 

Call:
logitr(data = yogurt, outcome = "choice", obsID = "obsID", pars = c("feat", 
    "brand"), scalePar = "price", randPars = c(feat = "n", brand = "n"), 
    panelID = "id", numMultiStarts = 10, numCores = 1)

Frequencies of alternatives:
       1        2        3        4 
0.402156 0.029436 0.229270 0.339138 

Summary Of Multistart Runs:
   Log Likelihood Iterations Exit Status
1       -1256.886        110           3
2       -1252.536         76           3
3       -1258.974         87           3
4       -1342.036        111           4
5       -1250.922        111           3
6       -1266.990         66           3
7       -1268.352         81           3
8       -1239.294         77           3
9       -1258.974         60           3
10      -1239.294         51           3

Use statusCodes() to view the meaning of each status code

Exit Status: 3, Optimization stopped because ftol_rel or ftol_abs was reached.
                                 
Model Type:           Mixed Logit
Model Space:   Willingness-to-Pay
Model Run:                8 of 10
Iterations:                    77
Elapsed Time:            0h:0m:3s
Algorithm:         NLOPT_LD_LBFGS
Weights Used?:              FALSE
Panel ID:                      id
Robust?                     FALSE

Model Coefficients: 
                  Estimate Std. Error  z-value  Pr(>|z|)    
scalePar          0.448563   0.039982  11.2191 < 2.2e-16 ***
feat              1.731133   0.491792   3.5201 0.0004315 ***
brandhiland     -14.223308   1.365310 -10.4176 < 2.2e-16 ***
brandweight      -8.172665   0.955928  -8.5495 < 2.2e-16 ***
brandyoplait      2.503597   0.407192   6.1484 7.825e-10 ***
sd_feat           1.266802   0.497472   2.5465 0.0108816 *  
sd_brandhiland   -7.114726   0.944233  -7.5349 4.885e-14 ***
sd_brandweight    9.130682   0.923411   9.8880 < 2.2e-16 ***
sd_brandyoplait   7.270250   0.752617   9.6600 < 2.2e-16 ***
---
Signif. codes:  0 '***' 0.001 '**' 0.01 '*' 0.05 '.' 0.1 ' ' 1
                                     
Log-Likelihood:         -1239.2939746
Null Log-Likelihood:    -3343.7419990
AIC:                     2496.5879491
BIC:                     2548.6819000
McFadden R2:                0.6293691
Adj McFadden R2:            0.6266775
Number of Observations:  2412.0000000

Summary of 10k Draws for Random Coefficients: 
             Min.     1st Qu.     Median       Mean   3rd Qu. Max.
feat         -Inf   0.8758265   1.730165   1.729359  2.584211  Inf
brandhiland  -Inf -19.0185364 -14.221025 -14.216712 -9.421990  Inf
brandweight  -Inf -14.3359189  -8.180155  -8.186754 -2.022659  Inf
brandyoplait -Inf  -2.4102743   2.491150   2.484915  7.394032  Inf
\end{CodeOutput}
\end{CodeChunk}

The summary shows the solution for iteration 8 out of the 10
multi-starts---the one with the largest log-likelihood value. Since the
\texttt{feat} and \texttt{brand} WTPs were both modeled as normally
distributed across the population, each of these covariates have two
parameters that describe the mean and standard deviation of a normal
distribution. The results again suggest that there is greater
heterogeneity for the ``Weight Watchers'' brand compared to ``Yoplait''
and ``Hiland'', which can be seen with its larger standard deviation
coefficient and wider WTP range in the distribution summary at the
bottom of the summary output. In fact, these results indicate that
although the mean WTP for the ``Weight Watchers'' brand is still higher
than that of the ``Hiland'' brand, the heterogeneity in WTP spans a much
wider range.

In this case, since the price parameter was modeled as a fixed
parameter, the WTP estimates from the preference space and those from
the WTP space model are nearly identical:

\begin{CodeChunk}
\begin{CodeInput}
R> wtpCompare(
+   model_pref = mxl_pref,
+   model_wtp  = mxl_wtp,
+   scalePar   = "price"
+ )
\end{CodeInput}
\begin{CodeOutput}
                         pref           wtp  difference
scalePar            0.4483378     0.4485634  0.00022557
feat                1.7330459     1.7311326 -0.00191335
brandhiland       -14.2021478   -14.2233082 -0.02116033
brandweight        -8.1828533    -8.1726652  0.01018805
brandyoplait        2.5036744     2.5035970 -0.00007740
sd_feat             1.2657757     1.2668022  0.00102646
sd_brandhiland     -7.0969787    -7.1147261 -0.01774740
sd_brandweight      9.1384873     9.1306816 -0.00780576
sd_brandyoplait     7.2741604     7.2702504 -0.00391001
logLik          -1239.2944250 -1239.2939746  0.00045043
\end{CodeOutput}
\end{CodeChunk}

If, however, the price parameter in either model were modeled as a
random parameter (which is controlled via the \texttt{randScale}
argument), the resulting WTP estimates could be substantially different.

\hypertarget{weighted-models}{%
\subsection{Weighted models}\label{weighted-models}}

Sometimes the modeler may wish to differentially weight individual
choice observations in model estimation. For example, if a particular
group was over- or under-represented in a sample relative to that of a
target population, the choice observations of that group could be
weighted such that they have a stronger or weaker contribution to the
log-likelihood in an attempt to balance the sample to match the
proportions of the target population.

The \texttt{cars\_us} data set that comes with the package includes a
\texttt{weights} column and is useful for illustrating how to estimate
weighted models. This data set contains 384 stated choice observations
from a conjoint survey of U.S. car buyers fielded online using Amazon
Mechanical Turk in 2012 and in person at the 2013 Pittsburgh Auto show
\citep{Helveston2015}. Participants were asked to select a vehicle from
a set of three alternatives, and each participant answered 15 choice
questions. The data set contains variables for different types of
vehicles (\texttt{"hev"}, \texttt{"phev10"}, \texttt{"phev20"},
\texttt{"phev40"}, \texttt{"bev75"}, \texttt{"bev100"},
\texttt{"bev150"}), different brands represented by the country of
origin (\texttt{"american"}, \texttt{"japanese"}, \texttt{"chinese"},
\texttt{"skorean"}), fast charging options (\texttt{"phevFastcharge"}
and \texttt{"bevFastcharge"}), price (\texttt{"price"}), operating cost
(\texttt{"opCost"}), and 0-60 mph acceleration time
(\texttt{"accelTime"}).

To compare the impact of the weights on the estimated parameters, an
unweighted and weighted model are estimated, both in the WTP space to
replicate the models estimated in Helveston et al.
\citeyearpar{Helveston2015}. In both models, the argument
\texttt{robust\ =\ TRUE} clusters the standard errors using the
\texttt{obsID} variable for clustering, which should be done for
weighted models. Standard errors can also be clustered at other levels
by specifying a \texttt{clusterID} variable. For example, a common
desired clustering is to cluster at the individual level in conjoint
studies where survey respondents answer multiple sequential choice
questions to account for potential correlations among these questions.
The unweighted model is estimated with the following code:

\begin{CodeChunk}
\begin{CodeInput}
R> set.seed(5678)
R> 
R> mnl_wtp_unweighted <- logitr(
+   data    = cars_us,
+   outcome = "choice",
+   obsID   = "obsnum",
+   pars = c(
+     "hev", "phev10", "phev20", "phev40", "bev75", "bev100", "bev150",
+     "american", "japanese", "chinese", "skorean", "phevFastcharge",
+     "bevFastcharge","opCost", "accelTime"),
+   scalePar = "price",
+   robust   = TRUE,
+   numMultiStarts = 10,
+   numCores = 1
+ )
\end{CodeInput}
\end{CodeChunk}

\begin{CodeChunk}
\begin{CodeInput}
R> summary(mnl_wtp_unweighted)
\end{CodeInput}
\begin{CodeOutput}
=================================================

Model estimated on: Tue Oct 04 11:59:52 2022 

Using logitr version: 0.8.0 

Call:
logitr(data = cars_us, outcome = "choice", obsID = "obsnum", 
    pars = c("hev", "phev10", "phev20", "phev40", "bev75", "bev100", 
        "bev150", "american", "japanese", "chinese", "skorean", 
        "phevFastcharge", "bevFastcharge", "opCost", "accelTime"), 
    scalePar = "price", robust = TRUE, numMultiStarts = 10, numCores = 1)

Frequencies of alternatives:
      1       2       3 
0.34323 0.33507 0.32170 

Summary Of Multistart Runs:
   Log Likelihood Iterations Exit Status
1       -4616.952         26           3
2       -4616.955         31           3
3       -4616.952         45           3
4       -4616.952         35           3
5       -4616.952         34           3
6       -4616.952         36           3
7       -4616.952         34           3
8       -4616.952         33           3
9       -4616.952         34           3
10      -4616.952         32           3

Use statusCodes() to view the meaning of each status code

Exit Status: 3, Optimization stopped because ftol_rel or ftol_abs was reached.
                                 
Model Type:     Multinomial Logit
Model Space:   Willingness-to-Pay
Model Run:                8 of 10
Iterations:                    33
Elapsed Time:          0h:0m:0.1s
Algorithm:         NLOPT_LD_LBFGS
Weights Used?:              FALSE
Cluster ID:                obsnum
Robust?                      TRUE

Model Coefficients: 
                  Estimate  Std. Error  z-value  Pr(>|z|)    
scalePar         0.0738787   0.0021929  33.6900 < 2.2e-16 ***
hev              0.8072448   0.9990581   0.8080 0.4190872    
phev10           1.1658652   1.0614987   1.0983 0.2720648    
phev20           1.6478081   1.0617443   1.5520 0.1206665    
phev40           2.5794026   1.0499274   2.4567 0.0140203 *  
bev75          -16.0458795   1.2541265 -12.7945 < 2.2e-16 ***
bev100         -13.0031631   1.2388544 -10.4961 < 2.2e-16 ***
bev150          -9.5733561   1.1641772  -8.2233 2.220e-16 ***
american         2.3442854   0.7979689   2.9378 0.0033053 ** 
japanese        -0.3747714   0.7998315  -0.4686 0.6393821    
chinese        -10.2685448   0.8859347 -11.5906 < 2.2e-16 ***
skorean         -6.0311955   0.8514340  -7.0836 1.405e-12 ***
phevFastcharge   2.8793913   0.8028804   3.5863 0.0003354 ***
bevFastcharge    2.9184681   0.9181323   3.1787 0.0014794 ** 
opCost          -1.6360487   0.0686313 -23.8382 < 2.2e-16 ***
accelTime       -1.6970364   0.1638091 -10.3598 < 2.2e-16 ***
---
Signif. codes:  0 '***' 0.001 '**' 0.01 '*' 0.05 '.' 0.1 ' ' 1
                                     
Log-Likelihood:         -4616.9517800
Null Log-Likelihood:    -6328.0067827
AIC:                     9265.9035600
BIC:                     9372.4426000
McFadden R2:                0.2703940
Adj McFadden R2:            0.2678655
Number of Observations:  5760.0000000
Number of Clusters       5760.0000000
\end{CodeOutput}
\end{CodeChunk}

The estimated WTP coefficients have units of \$1,000. The results
indicate large negative WTP values for the three full electric vehicle
types: -\$16,000 for \texttt{bev75}, -\$13,000 for \texttt{bev100}, and
-\$9,600 for \texttt{bev150} relative to conventional gasoline vehicles
(the numbers in each \texttt{bev} type indicate driving ranges in miles
on a full charge). There also appear to be strong brand preferences,
with WTP values ranging as much as -\$10,300 for Chinese brands to
\$2,300 for American brands relative to German brands.

To estimate a weighted model, the argument
\texttt{weights\ =\ "weights"} is added in the \texttt{logitr()}
function call. This sets the \texttt{weights} column in the
\texttt{cars\_us} data frame to be used to weight each choice
observation. The weights are specific to each individual survey
respondent and were calculated to account for over-sampling of younger
and less-wealthy car buyers \citep{Helveston2015}. In this example, the
weights have values ranging from \texttt{0.2} to \texttt{5}, meaning
some choice observations could have as much as 25 times the weight of
others in contributing to the log-likelihood.

\begin{CodeChunk}
\begin{CodeInput}
R> set.seed(5678)
R> 
R> mnl_wtp_weighted <- logitr(
+   data    = cars_us,
+   outcome = "choice",
+   obsID   = "obsnum",
+   pars = c(
+     "hev", "phev10", "phev20", "phev40", "bev75", "bev100", "bev150",
+     "american", "japanese", "chinese", "skorean", "phevFastcharge",
+     "bevFastcharge","opCost", "accelTime"),
+   scalePar = "price",
+   weights  = "weights",
+   robust   = TRUE,
+   numMultiStarts = 10,
+   numCores = 1
+ )
\end{CodeInput}
\end{CodeChunk}

\begin{CodeChunk}
\begin{CodeInput}
R> summary(mnl_wtp_weighted)
\end{CodeInput}
\begin{CodeOutput}
=================================================

Model estimated on: Tue Oct 04 11:59:55 2022 

Using logitr version: 0.8.0 

Call:
logitr(data = cars_us, outcome = "choice", obsID = "obsnum", 
    pars = c("hev", "phev10", "phev20", "phev40", "bev75", "bev100", 
        "bev150", "american", "japanese", "chinese", "skorean", 
        "phevFastcharge", "bevFastcharge", "opCost", "accelTime"), 
    scalePar = "price", weights = "weights", robust = TRUE, numMultiStarts = 10, 
    numCores = 1)

Frequencies of alternatives:
      1       2       3 
0.34323 0.33507 0.32170 

Summary Of Multistart Runs:
   Log Likelihood Iterations Exit Status
1       -3425.633         19           3
2       -3425.630         33           3
3       -3425.631         37           3
4       -3425.630         30           3
5       -3425.633         36           3
6       -3425.631         34           3
7       -3425.630         31           3
8       -3425.630         29           3
9       -3425.631         31           3
10      -3425.630         29           3

Use statusCodes() to view the meaning of each status code

Exit Status: 3, Optimization stopped because ftol_rel or ftol_abs was reached.
                                 
Model Type:     Multinomial Logit
Model Space:   Willingness-to-Pay
Model Run:               10 of 10
Iterations:                    29
Elapsed Time:         0h:0m:0.09s
Algorithm:         NLOPT_LD_LBFGS
Weights Used?:               TRUE
Cluster ID:                obsnum
Robust?                      TRUE

Model Coefficients: 
                  Estimate  Std. Error z-value  Pr(>|z|)    
scalePar         0.0522802   0.0040688 12.8489 < 2.2e-16 ***
hev             -1.1745214   2.9133014 -0.4032 0.6868318    
phev10           0.0275518   3.1280284  0.0088 0.9929723    
phev20           1.6949071   3.0997221  0.5468 0.5845208    
phev40           2.6494989   2.9851858  0.8875 0.3747834    
bev75          -20.1362768   3.6671641 -5.4910 3.997e-08 ***
bev100         -19.4967470   3.6256286 -5.3775 7.554e-08 ***
bev150         -13.6909374   3.4926845 -3.9199 8.859e-05 ***
american         8.1877347   2.4052979  3.4040 0.0006640 ***
japanese         0.9337835   2.3603628  0.3956 0.6923927    
chinese        -19.0068520   2.8539795 -6.6598 2.743e-11 ***
skorean         -9.5109238   2.5234809 -3.7690 0.0001639 ***
phevFastcharge   3.9438186   2.3624185  1.6694 0.0950384 .  
bevFastcharge    3.3428976   2.8087011  1.1902 0.2339704    
opCost          -1.5975429   0.1948476 -8.1989 2.220e-16 ***
accelTime       -1.1719313   0.4834735 -2.4240 0.0153513 *  
---
Signif. codes:  0 '***' 0.001 '**' 0.01 '*' 0.05 '.' 0.1 ' ' 1
                                     
Log-Likelihood:         -3425.6302862
Null Log-Likelihood:    -4360.5909275
AIC:                     6883.2605723
BIC:                     6989.7997000
McFadden R2:                0.2144115
Adj McFadden R2:            0.2107422
Number of Observations:  5760.0000000
Number of Clusters       5760.0000000
\end{CodeOutput}
\end{CodeChunk}

With both models estimated, it is helpful to directly compare the
estimated coefficients side-by-side:

\begin{CodeChunk}
\begin{CodeInput}
R> data.frame(
+   Unweighted = coef(mnl_wtp_unweighted),
+   Weighted   = coef(mnl_wtp_weighted)
+ )
\end{CodeInput}
\begin{CodeOutput}
                 Unweighted     Weighted
scalePar         0.07387865   0.05228019
hev              0.80724480  -1.17452143
phev10           1.16586524   0.02755184
phev20           1.64780809   1.69490706
phev40           2.57940264   2.64949894
bev75          -16.04587947 -20.13627677
bev100         -13.00316310 -19.49674699
bev150          -9.57335615 -13.69093743
american         2.34428544   8.18773467
japanese        -0.37477137   0.93378346
chinese        -10.26854481 -19.00685205
skorean         -6.03119552  -9.51092383
phevFastcharge   2.87939127   3.94381855
bevFastcharge    2.91846813   3.34289759
opCost          -1.63604869  -1.59754290
accelTime       -1.69703637  -1.17193131
\end{CodeOutput}
\end{CodeChunk}

From this comparison, it is clear that in the weighted model the
negative WTP for full electric vehicles is slightly larger than that in
the unweighted model, and the range of WTP for each brand also increased
in the weighted model. Nonetheless, all of the statistically significant
coefficients maintained the same sign and significance.

\hypertarget{predicting-probabilities}{%
\subsection{Predicting probabilities}\label{predicting-probabilities}}

Once a model has been estimated, it can be used to predict
probabilities, outcomes, or both for a set of alternatives using the
\texttt{predict()} method. Predictions can be made for any set of
alternatives so long as the columns in the alternatives correspond to
estimated coefficients in the model. By default, if no new data are
provided via the \texttt{newdata} argument, then predictions will be
made for the original data used to estimate the model.

Predictions can be made using both preference space and WTP space
models, as well as multinomial logit and mixed logit models. For mixed
logit models, heterogeneity is modeled by simulating draws from the
population estimates of the estimated model. In the example below, the
preference space MNL model from Section 5.5 (\texttt{mnl\_pref}) is used
to predict probabilities for the data used to estimate the model:

\begin{CodeChunk}
\begin{CodeInput}
R> probs <- predict(mnl_pref)
R> head(probs)
\end{CodeInput}
\begin{CodeOutput}
  obsID predicted_prob
1     1     0.41802407
2     1     0.02118240
3     1     0.23691737
4     1     0.32387615
5     2     0.26643822
6     2     0.02255486
\end{CodeOutput}
\end{CodeChunk}

The \texttt{predict()} method returns a data frame containing the
observation ID as well as the predicted probabilities. The original data
can also be returned in the data frame by setting
\texttt{returnData\ =\ TRUE}:

\begin{CodeChunk}
\begin{CodeInput}
R> probs <- predict(mnl_pref, returnData = TRUE)
R> head(probs)
\end{CodeInput}
\begin{CodeOutput}
  obsID predicted_prob price feat brandhiland brandweight brandyoplait choice
1     1     0.41802407   8.1    0           0           0            0      0
2     1     0.02118240   6.1    0           1           0            0      0
3     1     0.23691737   7.9    0           0           1            0      1
4     1     0.32387615  10.8    0           0           0            1      0
5     2     0.26643822   9.8    0           0           0            0      1
6     2     0.02255486   6.4    0           1           0            0      0
\end{CodeOutput}
\end{CodeChunk}

To make predictions for a new set of alternatives, use the
\texttt{newdata} argument. The example below makes predictions for just
two of the choice observations from the \texttt{yogurt} dataset:

\begin{CodeChunk}
\begin{CodeInput}
R> data <- subset(
+   yogurt, obsID %in% c(42, 13),
+   select = c('obsID', 'alt', 'price', 'feat', 'brand'))
R> 
R> probs_mnl_pref <- predict(
+   mnl_pref,
+   newdata = data,
+   obsID = "obsID",
+ )
R> 
R> probs_mnl_pref
\end{CodeInput}
\begin{CodeOutput}
  obsID predicted_prob
1    13     0.43685145
2    13     0.03312986
3    13     0.19155548
4    13     0.33846321
5    42     0.60764778
6    42     0.02602007
7    42     0.17803313
8    42     0.18829902
\end{CodeOutput}
\end{CodeChunk}

The \texttt{ci} argument can be used to obtain upper and lower bounds of
a confidence interval for predicted probabilities, which are estimated
using the Krinsky and Robb parametric bootstrapping method
\citep{Krinsky1986}. For example, a 95\% CI is obtained with
\texttt{ci\ =\ 0.95}:

\begin{CodeChunk}
\begin{CodeInput}
R> set.seed(5678)
R> 
R> probs_mnl_pref <- predict(
+   mnl_pref,
+   newdata = data,
+   obsID = "obsID",
+   ci = 0.95
+ )
R> 
R> probs_mnl_pref
\end{CodeInput}
\begin{CodeOutput}
  obsID predicted_prob predicted_prob_lower predicted_prob_upper
1    13     0.43685145           0.41629325           0.45825284
2    13     0.03312986           0.02608194           0.04111778
3    13     0.19155548           0.17572618           0.20815699
4    13     0.33846321           0.31873136           0.35843950
5    42     0.60764778           0.57346649           0.64162051
6    42     0.02602007           0.01872744           0.03598088
7    42     0.17803313           0.16138477           0.19504384
8    42     0.18829902           0.16787563           0.20791429
\end{CodeOutput}
\end{CodeChunk}

WTP space models can also be used to predict probabilities. In the
example below, the WTP space MNL model from Section 5.6
(\texttt{mnl\_wtp}) is used to predict probabilities for the
\texttt{data} object defined above:

\begin{CodeChunk}
\begin{CodeInput}
R> set.seed(5678)
R> 
R> probs_mnl_wtp <- predict(
+   mnl_wtp,
+   newdata = data,
+   obsID = "obsID",
+   ci = 0.95
+ )
R> 
R> probs_mnl_wtp
\end{CodeInput}
\begin{CodeOutput}
  obsID predicted_prob predicted_prob_lower predicted_prob_upper
1    13     0.43686141           0.41508522           0.45701654
2    13     0.03312947           0.02681696           0.04280579
3    13     0.19154829           0.17721984           0.20777595
4    13     0.33846083           0.31813769           0.35837404
5    42     0.60767120           0.57441503           0.64107061
6    42     0.02601800           0.01840558           0.03636293
7    42     0.17802363           0.16340562           0.19459951
8    42     0.18828717           0.16671672           0.20890971
\end{CodeOutput}
\end{CodeChunk}

\hypertarget{predicting-outcomes}{%
\subsection{Predicting outcomes}\label{predicting-outcomes}}

The \texttt{predict()} method can also be used to predict outcomes by
setting \texttt{type\ =\ "outcome"} (the default is \texttt{"prob"} for
predicting probabilities). In the examples below, outcomes are predicted
using the same preference space and WTP space models as in the previous
examples. The \texttt{returnData} argument is also set to \texttt{TRUE}
so that the predicted outcomes can be compared to the actual choices
made:

\begin{CodeChunk}
\begin{CodeInput}
R> set.seed(5678)
R> 
R> outcomes_pref <- predict(
+   mnl_pref,
+   type = "outcome",
+   returnData = TRUE
+ )
\end{CodeInput}
\end{CodeChunk}

\begin{CodeChunk}
\begin{CodeInput}
R> head(outcomes_pref)
\end{CodeInput}
\begin{CodeOutput}
  obsID predicted_outcome price feat brandhiland brandweight brandyoplait
1     1                 1   8.1    0           0           0            0
2     1                 0   6.1    0           1           0            0
3     1                 0   7.9    0           0           1            0
4     1                 0  10.8    0           0           0            1
5     2                 0   9.8    0           0           0            0
6     2                 0   6.4    0           1           0            0
  choice
1      0
2      0
3      1
4      0
5      1
6      0
\end{CodeOutput}
\end{CodeChunk}

\begin{CodeChunk}
\begin{CodeInput}
R> set.seed(5678)
R> 
R> outcomes_wtp <- predict(
+   mnl_wtp,
+   type = "outcome",
+   returnData = TRUE
+ )
\end{CodeInput}
\end{CodeChunk}

\begin{CodeChunk}
\begin{CodeInput}
R> head(outcomes_wtp)
\end{CodeInput}
\begin{CodeOutput}
  obsID predicted_outcome feat brandhiland brandweight brandyoplait scalePar
1     1                 1    0           0           0            0      8.1
2     1                 0    0           1           0            0      6.1
3     1                 0    0           0           1            0      7.9
4     1                 0    0           0           0            1     10.8
5     2                 0    0           0           0            0      9.8
6     2                 0    0           1           0            0      6.4
  choice
1      0
2      0
3      1
4      0
5      1
6      0
\end{CodeOutput}
\end{CodeChunk}

The accuracy of each model can be computed by dividing the number of
correctly predicted choices by the total number of choices:

\begin{CodeChunk}
\begin{CodeInput}
R> chosen_pref <- subset(outcomes_pref, choice == 1)
R> chosen_pref$correct <- chosen_pref$choice == chosen_pref$predicted_outcome
R> accuracy_pref <- sum(chosen_pref$correct) / nrow(chosen_pref)
R> accuracy_pref
\end{CodeInput}
\begin{CodeOutput}
[1] 0.3706468
\end{CodeOutput}
\begin{CodeInput}
R> chosen_wtp <- subset(outcomes_wtp, choice == 1)
R> chosen_wtp$correct <- chosen_wtp$choice == chosen_wtp$predicted_outcome
R> accuracy_wtp <- sum(chosen_wtp$correct) / nrow(chosen_wtp)
R> accuracy_wtp
\end{CodeInput}
\begin{CodeOutput}
[1] 0.3706468
\end{CodeOutput}
\end{CodeChunk}

These results show that both models correctly predicted choice for
approximately 37\% of the observations in the \texttt{yogurt} data
frame, which is significantly better than random (25\%).

\hypertarget{additional-options}{%
\subsection{Additional options}\label{additional-options}}

The \texttt{logitr()} function contains many other arguments for
controlling different aspects of the model specification and estimation
procedure. For example, the estimated solution for mixed logit models
can sometimes be sensitive to the resolution of the simulated random
parameter distributions. By default, 50 Halton draws are used, but this
can be increased using the \texttt{numDraws} argument. The user can also
use Sobol draws by specifying \texttt{drawType\ =\ "sobol"} (defaults to
\texttt{"halton"}), which is recommended in models with a larger number
of random parameters \citep{Czajkowski2019}.

Details of the optimization procedure can be controlled via the
\texttt{options} argument, which must be a named list of control
options. For example, the optimization tolerance levels can be
controlled by changing the values for \texttt{xtol\_rel},
\texttt{xtol\_abs}, \texttt{ftol\_rel}, and \texttt{ftol\_abs}. The
function \texttt{nloptr::nloptr.print.options()} prints details on these
control settings to the console.

Finally, it can be helpful to provide a custom set of starting values
for models that have trouble converging. For WTP space models in
particular, one strategy is to first compute the WTP from a preference
space model and then use those results as the starting values. For
example, using the \texttt{mnl\_pref} model estimated in Section 5.5, we
can first compute the corresponding WTP using the \texttt{wtp()}
function:

\begin{CodeChunk}
\begin{CodeInput}
R> wtp_est <- wtp(mnl_pref, scalePar = "price")$Estimate
\end{CodeInput}
\end{CodeChunk}

The computed \texttt{wtp\_est} vector can then be passed to the
\texttt{startVals} argument for the \texttt{logitr()} function. If a
multi-start is used, the user-provided starting values will only be used
for the first iteration.

\begin{CodeChunk}
\begin{CodeInput}
R> set.seed(5678)
R> 
R> mnl_wtp2 <- logitr(
+   data      = yogurt,
+   outcome   = 'choice',
+   obsID     = 'obsID',
+   pars      = c('feat', 'brand'),
+   scalePar  = 'price',
+   startVals = wtp_est,
+   numMultiStarts = 10,
+   numCores = 1
+ )
\end{CodeInput}
\end{CodeChunk}

\newpage

\hypertarget{limitations-of-wtp-space-models}{%
\section{Limitations of WTP space
models}\label{limitations-of-wtp-space-models}}

Although package \pkg{logitr} was designed to simplify the estimation of
WTP space models, these models do have several important limitations.
First, since only one scale parameter is estimated in a WTP space model,
the true WTP could be over- or under-estimated if there are multiple
latent classes in the sample that each have different sensitivities to
price. In contrast, in preference space models interactions can be used
to estimate these differences in price sensitivities for different
groups in a given sample, which could then be used to compute WTP for
each group. While one could estimate separate WTP space models on each
latent class to account for differences in price sensitivities by group,
this requires that every model parameter be separately estimated across
each group, which is a stricter assumption. A latent class
implementation of WTP space models has not yet been explored.

In addition, given the non-linear utility specification of WTP space
models, these models can often diverge during estimation and can be
highly sensitive to starting parameters. Models in which the scale
parameter is modeled as a random parameter in particular tend to diverge
more often during estimation. For example, if the scale parameter is
assumed to be log-normally distributed to force positivity (which can be
done by setting \texttt{randScale\ =\ "ln"} in the \texttt{logitr()}
function), the model may not consistently converge on a solution as the
draws from this distribution can sometimes have extremely large values
that could have dramatic effects on the optimization search.

Even so, \{logitr\} tends to perform better and converge more often
compared to many other packages that support WTP space models. By using
the parallelized multi-start optimization loop, the package can
efficiently search for different local minima from different random
starting points when minimizing the negative log-likelihood, improving
the chances of converging to a solution for at least some of the
multi-start iterations. For more details, see the package vignette
titled ``WTP space convergence issues in other packages''.

Finally, WTP space models can be computationally expensive. Since WTP
space models have a non-convex log-likelihood function, there is no
guaranteed that any one iteration of the optimization algorithm will
reach a global solution. As a result, it is recommended that a
multi-start optimization loop always be used for WTP space models, which
increases computation time. In practice, it may be helpful to initially
use a relatively small number of multi-start iterations (e.g., 10) to
ensure that the optimization is converging in most iterations before
using a larger number of iterations to conduct a broader search. The
\texttt{startValBounds} argument, which is set to \texttt{c(-1,\ 1)} by
default, can be used to specify the lower and upper boundaries of the
random starting parameters used in each iteration of a multi-start loop.

\hypertarget{conclusions}{%
\section{Conclusions}\label{conclusions}}

Package \pkg{logitr} implements the maximum likelihood estimation of
multinomial logit and mixed logit models with unobserved heterogeneity
across individuals, which is modeled by allowing parameters to vary
randomly over individuals according to a chosen distribution.
Distinguishing features include fast estimation speeds and support for
utility models that can be specified using either a ``preference space''
or ``WTP space'' parameterization, allowing for the direct estimation of
marginal WTP. This offers several advantages over the typical procedure
of computing WTP using the estimated parameters of a preference space
model, including greater control over how WTP is assumed to be
distributed across the population.

While \pkg{logitr} is less general in scope than other similar packages
in terms of the variety of supported models, it is considerably faster
and offers other functionality that is particularly useful for
estimating WTP space and mixed logit models. For example, a parallelized
multi-start optimization loop offers a convenient interface for
searching the solution space for different local minima when estimating
models with non-convex log-likelihood functions (i.e., WTP space and
mixed logit models). In addition, although the user interface departs
from the popular \texttt{formula} input, it is more uniform and
streamlined for estimating models with preference or WTP utility
parameterizations.

The package could be further improved by adding support for other
features, such as individual-level parameter estimates, and other random
parameter distributions, such as the triangle distribution and the
exponential distribution. The package source code and documentation can
be found at \url{https://github.com/jhelvy/logitr}.

\hypertarget{acknowledgements}{%
\section*{Acknowledgements}\label{acknowledgements}}
\addcontentsline{toc}{section}{Acknowledgements}

I would like to express my gratitude to Connor Forsythe for contributing
to the package by adding support for clustering errors, Elea Feit for
providing critical feedback in early drafts of this manuscript, Jeremy
Michalek for introducing me to choice modeling, Kenneth Train for his
incredibly well-written and accessible texts, and the JSS reviewers
whose comments and suggestions substantially improved the package and
this manuscript.

\renewcommand\refname{References}
\bibliography{jss4524.bib}

\end{document}